\documentclass[aps,nofootinbib,prd]{revtex4}
\usepackage{graphicx}
\usepackage{epsfig}
\usepackage{amsmath,amssymb}

 \newcommand{\be}{\begin{eqnarray}}
 \newcommand{\ee}{\end{eqnarray}}
 \newcommand{\al}{\alpha}

 \newcommand{\La}{\Lambda}
 \newcommand{\Om}{\Omega}
 \newcommand{\dd}{\partial}

\begin{document}
\title{DARK ENERGY AND MODIFIED GRAVITY }

\author{Ruth Durrer$^1$ and Roy Maartens$^2$}

\affiliation{$^1$D\'epartment de Physique Th\'eorique,
              Universit\'e de Gen\`eve, 1211
              Gen\`eve 4, Switzerland\\
             $^2$Institute of Cosmology \& Gravitation,
              University of Portsmouth, Portsmouth PO1 2EG, UK}

\begin{abstract}

Explanations of the late-time cosmic acceleration within the
framework of general relativity are plagued by difficulties.
General relativistic models are mostly based on a dark energy
field with fine-tuned, unnatural properties. There is a great
variety of models, but all share one feature in common -- an
inability to account for the gravitational properties of the
vacuum energy, and a failure to solve the so-called coincidence
problem. Two broad alternatives to dark energy have emerged as
candidate models: these typically address only the coincidence
problem and not the vacuum energy problem. The first is based on
general relativity and attempts to describe the acceleration as an
effect of inhomogeneity in the universe. If this alternative could
be shown to work, then it would provide a dramatic resolution of
the coincidence problem; however, a convincing demonstration of
viability has not yet emerged. The second alternative is based on
infra-red modifications to general relativity, leading to a
weakening of gravity on the largest scales and thus to
acceleration. Most examples investigated so far are scalar-tensor
or brane-world models, and we focus on the simplest candidates of
each type: $f(R)$ models and DGP models respectively. Both of
these provide a new angle on the problem, but they also face
serious difficulties. However, investigation of these models does
lead to valuable insights into the properties of gravity and
structure formation, and it also leads to new strategies for
testing the validity of General Relativity itself on cosmological
scales.
\end{abstract}

\maketitle

\section{INTRODUCTION}

\begin{figure*}
\begin{center}
\includegraphics[height=3.25in,width=3.in]{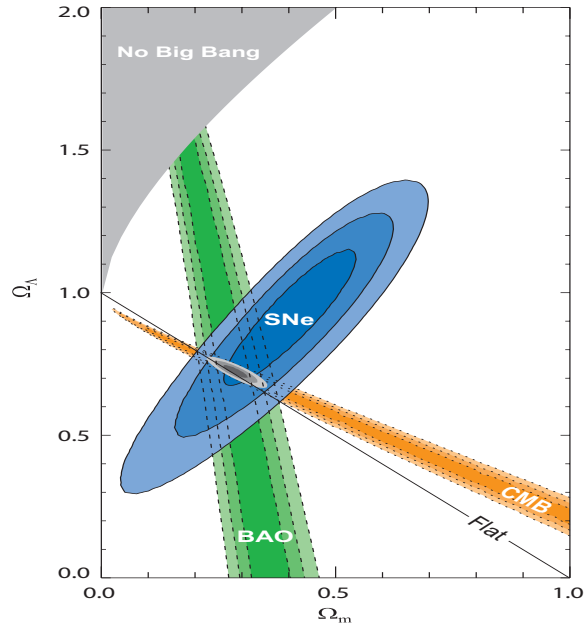}
\end{center}
\caption{Observational constraints in the
$(\Omega_{m},\Omega_\Lambda)$ plane: joint constraints from
supernovae (SNe), baryon acoustic oscillations (BAO) and CMB
(from~\cite{Kowalski:2008ez}). } \label{sn}
\end{figure*}

The current ``standard model" of cosmology is the inflationary
cold dark matter model with cosmological constant $\Lambda$,
usually called LCDM, which is based on general relativity and
particle physics (i.e., the Standard Model and its minimal
supersymmetric extensions). This model provides an excellent fit
to the wealth of high-precision observational data, on the basis
of a remarkably small number of cosmological
parameters~\cite{data}. In particular, independent data sets from
cosmic microwave background (CMB) anisotropies, galaxy surveys and
supernova luminosities, lead to a consistent set of best-fit model
parameters (see Fig.~\ref{sn}) -- which represents a triumph for
LCDM.

The standard model is remarkably successful, but we know that its
theoretical foundation, general relativity, breaks down at high
enough energies, usually taken to be at the Planck scale,
\begin{equation}
E \gtrsim M_p \sim 10^{16}\,\mbox{TeV}\,.
\end{equation}
The LCDM model can only provide limited insight into the very
early universe. Indeed, the crucial role played by inflation
belies the fact that inflation remains an effective theory without
yet a basis in fundamental theory. A quantum gravity theory will
be able to probe higher energies and earlier times, and should
provide a consistent basis for inflation, or an alternative that
replaces inflation within the standard cosmological model.

An even bigger theoretical problem than inflation is that of the
late-time acceleration in the expansion of the
universe~\cite{de,grg}. In terms of the fundamental energy density
parameters, the data indicates that the present cosmic energy
budget is given by (see Fig.~\ref{sn})
 \be
\label{olom} \Omega_\Lambda \equiv {\Lambda \over 3H_0^2}\approx
0.75\,,~~ \Omega_{m} \equiv {8\pi G\rho_{m0} \over 3H_0^2}\approx
0.25\,, ~~ \Omega_{K} \equiv {-K\over a_0^2H_0^2} \approx 0 \,, ~~
\Omega_r \equiv {8\pi G\rho_{r0} \over 3H_0^2}\approx 8\times
10^{-5}\,.
 \ee
Here $H_0=100h$km$($s Mpc$)^{-1}$ is the present value of the
Hubble parameter, $\La$ is the cosmological constant, $K$ is
spatial curvature, $\rho_{m0}$ is the present matter density
and $\rho_{r0}$ is the present radiation density. Newton's
constant is related to the Planck mass by $G=M_p^{-2}$ (we
use units where the speed of light, $c=1$ and Planck's constant
$\hbar=1$).

The Friedmann equation governs the evolution of the
scale factor $a(t)$. It is given by
\begin{eqnarray}
\left({\dot a\over a}\right)^2 \equiv H^2 &=& {8\pi G\over
3}(\rho_m+\rho_r)+{\Lambda \over 3}-{K \over a^2} \nonumber\\ &=&
H_0^2\left[\Omega_m(1+z)^3+ \Omega_r(1+z)^4 +\Omega_\Lambda+
\Omega_K(1+z)^2 \right], \label{h}
\end{eqnarray}
The scale factor, which is related to the
cosmological redshift by $z=a^{-1}-1$. (We normalize the present
scale factor to $a_0=1$.) Together with the energy conservation
equation this implies
\begin{equation}
{\ddot a \over a} = -{4\pi G \over 3} \left(\rho_m+2\rho_r\right)
+ {\Lambda \over 3} \,. \label{acc}
\end{equation}
The observations, which together with Eq.~(\ref{h}) lead to the
values given in Eq.~(\ref{olom}), produce via Eq.~(\ref{acc}) the
dramatic conclusion that the universe is currently accelerating,
 \be
\ddot a_0=H_0^2\left(\Omega_\Lambda-{1 \over 2} \Omega_m-\Omega_r
\right)>0\,.
 \ee

This conclusion holds only if the universe is (nearly) homogeneous
and isotropic, i.e., a Friedmann-Lema\^\i tre model. In this case
the distance to a given redshift $z$, and the time elapsed since
that redshift, are tightly related via the only free function of
this geometry, $a(t)$. If the universe instead is isotropic around
us but not homogeneous, i.e., if it resembles a
Tolman-Bondi--Lema\^\i tre solution with our galaxy cluster at the
centre, then this tight relation between distance and time for a
given redshift would be lost and present data would not
necessarily imply acceleration -- or the data may imply
acceleration without dark energy. This remains a controversial and
unresolved issue (see e.g.~\cite{kari}).

Of course isotropy without homogeneity violates the Copernican
Principle as it puts us in the centre of the Universe. However, it
has to be stressed that up to now observations of homogeneity are
very limited, unlike isotropy, which is firmly established.
Homogeneity is usually inferred from isotropy together with the
Copernican principle. With future data, it will in principle be
possible to distinguish observationally an isotropic but
inhomogeneous universe from an isotropic and homogeneous universe
(see e.g.~\cite{Goodman}). Testing the Copernican Principle is a
crucial aspect of testing the standard cosmological model. But in
the following, we will assume that the Copernican Principle
applies.

The data also indicate that the universe is currently (nearly)
spatially flat,
 \be
|\Omega_K|\ll 1\,.
 \ee
It is common to assume that this implies $K=0$ and to use
inflation as a motivation. However, inflation does {\em not} imply
$K=0$, but only $\Omega_K\to 0$. Even if this distinction may be
negligible in the present universe,  a nonzero curvature can have
significant implications for the onset of inflation(see
e.g.~\cite{ellmaa}). In fact, if the present curvature is small but
non-vanishing, neglecting it in the analysis of Supernova data can
sometimes induce surprisingly large errors~\cite{brucechris}.

The simplest way to explain acceleration is probably a
cosmological constant, i.e., the LCDM model. Even though the
cosmological constant can be considered as simply an additional
gravitational constant (in addition to Newton's constant), it
enters the Einstein equations in exactly the
same way as a contribution from the vacuum energy, i.e., via a
Lorentz-invariant energy-momentum tensor $T^{\rm vac}_{\mu\nu}=
-(\Lambda/8\pi G) g_{\mu\nu}$. The only observable signature of
both a cosmological constant and vacuum energy is their effect on
spacetime -- and so a vacuum energy and a classical cosmological
constant cannot be distinguished by observation. Therefore the
`classical' notion of the cosmological constant is effectively
physically indistinguishable from quantum vacuum energy.

Even though the absolute value of vacuum energy cannot be
calculated within quantum field theory, {\em changes} in the
vacuum energy (e.g. during a phase transition) can be calculated,
and they do have a physical effect -- for example, on the energy
levels of atoms (Lamb shift), which is well known and well
measured. Furthermore, differences of vacuum energy in different
locations, e.g., between or on one side of two large metallic
plates, have been calculated and their effect, the Casimir force,
is well measured~\cite{casi}. Hence, there is no doubt about the
reality of vacuum energy. For a field theory with cutoff energy
scale $E$, the vacuum energy density scales with the cutoff as
$\rho_{\rm vac} \sim E^4$, corresponding to a cosmological
constant $\La_{\rm vac} =8\pi G\rho_{\rm vac}$.  If $E=M_{p}$,
this yields a na\"\i ve contribution to the `cosmological constant' of
about $\La_{\rm vac} \sim 10^{38}$GeV$^2$, whereas the measured
effective cosmological constant is the sum of the `bare'
cosmological constant and the contribution from the cutoff scale,
 \be
\La_\mathrm{eff}= \La_{\rm vac}+\La \simeq
10^{-83}\,\mathrm{GeV}^2~. \label{v1}
 \ee
Hence a cancellation of about 120 orders of magnitude is required.
This is called the {\em fine-tuning} or {\em size } problem of
dark energy: a cancellation is needed to arrive at a result which
is many orders of magnitude smaller than each of the
terms.\footnote{In quantum field theory we actually have to add to
the
  cut-off term $\La_{\rm vac} \simeq E_c^4/M_{pl}^2$ the unmeasurable
  `bare'
  cosmological constant. In this sense, the cosmological constant
  problem is a fine tuning
  between the unobservable `bare' cosmological constant and the term
  coming from the cut-off scale.}
It is possible that the quantum vacuum energy is much smaller than
the Planck scale. But even if we set it to the lowest possible
SUSY scale, $E_{\rm susy}\sim 1$TeV, arguing that at higher
energies vacuum energy exactly cancels due to supersymmetry, the
required cancellation is still about 60 orders of magnitude.

A reasonable attitude towards this open problem is the hope that
quantum gravity will explain this cancellation. But then it is
much more likely that we shall obtain directly $\La_{\rm
vac}+\La=0$ and not $\La_{\rm vac}+\La \simeq 24\pi G\rho_m(t_0)$.
This unexpected observational result leads to a second problem,
{\em the coincidence problem}: given that
 \be
\rho_\Lambda = {\Lambda_\mathrm{eff} \over 8\pi
G}=\,\mbox{constant}\,,~\mbox{ while }~\rho_m \propto (1+z)^3\,,
 \ee
why is  $\rho_\Lambda$ of the order of the {\em present} matter
density $\rho_m(t_0)$? It was completely negligible in most of the
past and will entirely dominate in the future.

These problems prompted cosmologists to look for other explanation
of the observed accelerated expansion.
Instead of a cosmological constant, one may introduce a
scalar field or some other contribution to the energy-momentum
tensor which has an equation of state $w<-1/3$. Such a component
is called `dark energy'. So far, no consistent model of dark
energy has been proposed which can yield a convincing or natural
explanation of either of these problems (see, e.g.~\cite{linder}).

Alternatively, it is possible that there is no dark energy field,
but instead the late-time acceleration is a signal of a {\em
gravitational} effect. Within the framework of general relativity,
this requires that the impact of inhomogeneities somehow acts to
produce acceleration, or the appearance of acceleration (within a
Friedman-Lema\^{i}tre interpretation). A non-Copernican
possibility is the Tolman-Bondi--Lema\^\i tre model~\cite{kari}.
Another (Copernican) possibility is that the `backreaction' of
inhomogeneities on the background, treated via nonlinear
averaging, produces effective acceleration~\cite{thomas}.

A more radical version is the `dark gravity' approach, the idea
that gravity itself is weakened on large-scales, i.e., that there
is an ``infrared'' modification to general relativity that
accounts for the late-time acceleration. The classes of modified
gravity models which have been most widely investigated are
scalar-tensor models~\cite{CapFran} and brane-world
models~\cite{koyama}.

Schematically, we are modifying the geometric side of the field
equations,
 \be
G_{\mu\nu}+G^{\rm dark}_{\mu\nu} = 8\pi G T_{\mu\nu}\,,
\label{mod}
 \ee
rather than the matter side,
 \be
G_{\mu\nu} = 8\pi G \left(T_{\mu\nu}+ T^{\rm dark}_{\mu\nu}
\right)\,,
 \ee
as in the general relativity approach. Modified gravity represents
an intriguing possibility for resolving the theoretical crisis
posed by late-time acceleration. However, it turns out to be
extremely difficult to modify general relativity at low energies
in cosmology, without violating observational constraints -- from
cosmological and solar system data, or without introducing ghosts
and other instabilities into the theory. Up to now, there is no
convincing alternative to the general relativity dark energy
models -- which themselves are not convincing.

The plan of the remainder of this chapter is as follows. In
Section~2 we discuss constraints which one may formulate for a
dark energy or modified gravity theory from basic theoretical
requirements. In Section~3 we briefly discuss models that address
the dark energy problem within general relativity. In Section~4 we
present modified gravity models. In Section~5 we conclude.
This article is based on a previous review published in~\cite{RoyRuthold}.

\section{CONSTRAINING EFFECTIVE THEORIES}
\label{s:const}

Theories of both dark matter and dark energy often have very
unusual Lagrangians that cannot be quantized in the usual way,
e.g. because they have non-standard kinetic terms. We then simply
call them `effective low energy theories' of some unspecified high
energy theory which we do not elaborate. In this section, we want
to point out a few properties which we nevertheless can require of
low energy effective theories.  We first enumerate the properties
which we can require from a good basic physical theory at the
classical and at the quantum level. We then discuss which of these
requirements are inherited by low energy effective descriptions.

\subsection{{\bf FUNDAMENTAL PHYSICAL THEORIES}}
\label{sec:fund}

Here we give a minimal list of properties which we require from a
fundamental physical theory. Of course, all the points enumerated
below are open for discussion, but at least we should be aware of
what we lose when we let go of them.

In our list we start with very basic requirements which become
more strict as we go on. Even though some theorists would be able
to live without one or several of the criteria discussed here, we
think they are all very well founded. Furthermore, all known
current physical theories, including string- and M-theory, do
respect them.

\setcounter{enumi}{-1}
\begin{enumerate}

\item {\bf A physical theory allows a mathematical description}\\
This is the basic idea of theoretical physics. \vspace{0.1cm}

\item {\bf A fundamental physical theory allows a Lagrangian formulation}\\
This
requirement is of course much stronger than the previous one. But
it has been extremely successful in the past and was the guiding
principle for the entire development of quantum field theory and
string theory in the 20th century.

Some `varying speed of light theories' without Lagrangian
formulation leave us more or less free to specify the evolution of
the speed of light during the expansion history of the universe.
However, if we introduce a Lagrangian formulation, we realize that
most of these theories are simply some variant of scalar-tensor
theories of gravity, which are well defined and have
been studied in great detail.

If we want to keep deep physical insights like N\"other's theorem,
which relates symmetries to conservation laws, we need to require
a Lagrangian formulation for a physical theory. A basic ingredient
of a Lagrangian physical theory is that every physical degree of
freedom has a kinetic term which consists (usually) of first order
time derivatives and may also have a `potential term' which does
not involve derivatives. In the Lagrangian formulation of a
fundamental physical theory, we do not allow for external,
arbitrarily given functions. Every function has to be a degree of
freedom of the theory so that its evolution is determined
self-consistently via the Lagrangian equations of motion, which
are of first or second order. It is possible that the Lagrangian
contains also higher than first order derivatives, but such
theories are strongly constrained by the problem of ghosts which
we mention below, and by the fact that the corresponding equations
of motion are usually described by an unbounded Hamiltonian, i.e.
the system is unstable (Ostrogradski's
theorem~\cite{Ostro,Woody}). \vspace{0.1cm}

\item {\bf Lorentz invariance}\\
We also want to require that the theory be Lorentz invariant. Note
that this requirement is much stronger than demanding simply
`covariance'. It requires that there be no `absolute element' in
the theory apart from true constants. Lorentz covariance can
always be achieved by rewriting the equations. As an example,
consider a Lagrangian given in flat space by
$(\dd_t\phi)^2-(\dd_x\phi)^2$. This is clearly not Lorentz
invariant. However, we can trivially write this term in the
covariant form $\alpha^{\mu\nu}\dd_\nu\dd_\mu\phi$, by setting
$(\alpha^{\mu\nu}) = {\rm diag}(1,-1,0,0)$. Something like this
should of course not be allowed in a fundamental theory. A term of
the form $\alpha^{\mu\nu}\dd_\nu\dd_\mu\phi$ is only allowed if
$\alpha^{\mu\nu}$ is itself a dynamical field of the theory. This
is what we mean by requiring that the theory is not allowed to
contain `absolute elements', i.e. it is Lorentz invariant and not
simply covariant.  \vspace{0.1cm}

\item {\bf Ghosts }\\
Ghosts are fields whose kinetic term has the wrong sign. Such a
field, instead of slowing down when it climbs up a potential, is
speeding up. This unstable situation leads to severe problems when
we want to quantize it, and it is generally accepted that one
cannot make sense of such a theory, at least not at the quantum level.
This is not surprising, since quantization usually is understood
as defining excitations above some ground state, and a theory with
a ghost has no well defined ground state. Its kinetic energy has
the wrong sign and the larger $\dot\phi^2$ is, the lower is the
energy. \vspace{0.1cm}

 \item {\bf Tachyons }\\
These are degrees of freedom that have a negative mass squared,
$m^2<0$. Using again the simple scalar field example, this means
that the second derivative of the potential about the `vacuum
value' ($\phi = 0$ with $\dd_\phi V(0)=0$) is negative,
$\dd^2_\phi V(0)< 0$. In general, this need not mean that the
theory makes no sense, but rather that $\phi=0$ is a bad choice
for expanding around, since it is a maximum rather than a minimum
of the potential and therefore an unstable equilibrium.

This means also that the theory cannot be quantized around the
classical solution $\phi=0$, but it may become a good quantum
theory by a simple shift, $\phi\rightarrow \phi -\phi_0$, where
$\phi_0$ is the minimum of the potential. If the potential of a
fundamental scalar field has no minimum but only a maximum, the
situation is more severe. Then the theory is truly unstable.
\vspace{0.1cm}

The last two problems, together with the Ostrogradski instability
that appears in theories with higher derivatives, can be
summarized in the requirement that a meaningful theory needs to
have an energy functional which is bounded from below.
\vspace{0.1cm}

\item {\bf Superluminal motion and causality}\\
A fundamental physical theory which does respect Lorentz
invariance must not allow for superluminal motions. If this
condition is not satisfied, we can construct closed curves along
which a signal can propagate~\cite{closed}. (See
Fig.~\ref{f:closed}.)

\begin{figure}[ht]
\centerline{ \epsfig{file=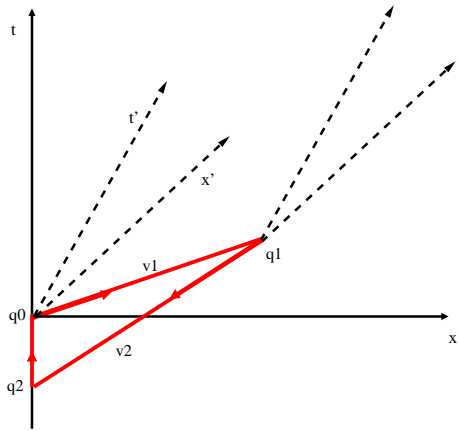, width=6cm}}
\caption{\label{f:closed} We assume a signal that can propagate
at speeds $v_1\,,\, v_2>1$. The frame R' with coordinates $(t',x')$
moves with speed $v<1$ in the x-direction. The speed $v$ is
chosen such that both, $v_1\,,\, v_2>1/v$. A signal is sent with
velocity $v_1$ from $q_0$ to
  $q_1$ in the frame R. Since $v_1>1/v$, this signal travels backward
  in time with respect to frame $R'$. Then a signal is sent with speed
  $v_2$ from $q_1$ to $q_2$. Since $|v_2|>1/v$, this signal, which is sent
  forward in time in frame $R'$, travels backward in time with respect
  to $R$ and can arrive at an event $q_2$ with $t_2<0$.
  The loop generated in this way is not `causal' since both the
trajectory from $q_0$ to $q_1$ and the one from $q_1$ to $q_2$ are
spacelike. So we cannot speak of the formation of closed causal
loops, but it is nevertheless a closed loop along which a signal
can propagate and which therefore enables the construction of a
time machine, leading to the usual problems with causality and
entropy. (From~\cite{closed}.)}
\end{figure}

At first sight one might think that a Lorentz invariant Lagrangian
will automatically forbid superluminal motions. But the situation
is not so simple. Generic Lorentz invariant higher spin theories,
$s\ge 1$, lead to superluminal motion~\cite{velo}. While the
equations are manifestly Lorentz invariant, their characteristics
in general do not coincide with the light cone and can very well
be spacelike. There are exceptions, among which are Yang Mills
theories for spin 1 and the linearized Einstein equations for spin
2.

One may object to this restriction, on the grounds that general
relativity, which is certainly a theory that is acceptable (at
least at the classical level), can lead to closed causal curves,
even though it does not admit superluminal
motion~\cite{Thorne,Gott,Ori,Bonnor}.\vspace{0.1cm}

The situation is somewhat different if superluminal motion is only
possible in a background which breaks Lorentz-invariance. Then one
has in principle a preferred frame and one can specify that
perturbations should always propagate with the Green's function
that corresponds to the retarded Green's function in this
frame~\cite{muki}. Nevertheless, one has to accept that there will
be boosted frames relative to which the Cauchy problem for the
superluminal modes is not well defined. The physics experienced by
an observer in such a frame is most unusual (to say the least).

Causality of a theory is intimately related to  the analyticity
properties of the $S$-matrix of scattering, without which
perturbative quantum theory does not make sense. Furthermore, we
require the $S$ matrix to be unitary. Important consequences of
these basic requirements are the Kramers Kronig dispersion
relations, which are a result of the analyticity properties and
hence of causality, and the optical theorem, which is a result of
unitarity. The analyticity properties have many further important
consequences, such as the Froissart bound, which implies that the
total cross section converges at high energy~\cite{KKetal}.

\end{enumerate}

\subsection{{\bf LOW ENERGY EFFECTIVE THEORIES}}
\label{sec:low}

The concept of low energy effective theories is extremely useful
in physics. As one of the most prominent examples, consider
superconductivity. It would be impossible to describe this
phenomenon by using full quantum electrodynamics with a typical
energy scale of MeV, where the energy scale of superconductivity
is milli-eV and less. However, many aspects of superconductivity
can be successfully described with the Ginzburg-Landau  theory of
a complex scalar field. Microscopically, this scalar field is to
be identified with a Cooper pair of two electrons, but this is
irrelevant for many aspects of superconductivity.

Another example is  weak interaction and four-Fermi theory. The
latter is a good approximation to weak interactions at energy
scales far below the $Z$-boson mass. Most physicists also regard
the standard model of particle physics as a low energy effective
theory which is valid below some high energy scale beyond which
new degrees of freedom become relevant, be this supersymmetry, GUT
or string theory.

We now want to investigate which of the properties in the previous
subsection may be lost if we `integrate out' high energy
excitations and consider only processes which take place at
energies below some cutoff scale $E_c$. We cannot completely
ignore all particles with masses above $E_c$, since in the low
energy quantum theory they can still be produced `virtually',
i.e., for a time shorter than $1/E_c$. This is not relevant for
the initial and final states of a scattering process, but plays a
role in the interaction.

Coming back to our list in the previous subsection, we certainly
want to keep the first point -- a mathematical description. But
the Lagrangian formulation will also survive if we proceed in a
consistent way by simply integrating out the high energy degrees
of freedom.

What about higher order derivatives in the Lagrangian? The problem
is that, in general, there is no Hamiltonian that is bounded from
below if the Lagrangian contains higher derivatives, i.e. the
system is unstable~\cite{Woody}. Of course it is possible to find
well behaved solutions of this system, since for a given solution
energy is conserved. But as soon as the system is interacting,
with other degrees of freedom, it will lower its energy and
produce more and more modes of these other degrees of freedom.
This is especially serious when one quantizes the system.
The vacuum is exponentially unstable to simultaneous production
of modes of positive and negative energy. Of course one cannot
simply `cut away' the negative energy solutions without violating
unitarity. And even if the theory under consideration is only a
low energy effective theory, it should at least be `unitary at
low energy'. Introducing even higher derivatives only worsens
the situation, since the Hamiltonian acquires more unstable directions.

For this argument, it does not matter whether the degrees of
freedom we are discussing are fundamental or only low energy
effective degrees of freedom. Even if we modify the Hamiltonian at
high energies, the instability, which is a {\em low energy
problem}, will not disappear. There are only two ways out of the
Ostrogradski instability: Firstly, if the necessary condition that
the lagrangian be non-degenerate is not satisfied. The second
possibility is via constraints, whereby one might be able to
eliminate the unstable directions. In practice, this has to be
studied on a case by case basis.
An important example for the dark energy problem, which avoids
the Ostrogradski instability via constraints, are modified
gravity Lagrangians of the form $f(R)$, discussed below.

If  the Ostradgradski theorem does not apply, we have still no
guarantee that the theory has no ghosts or that the potential
energy is bounded from below (no `serious' tachyon). The
limitation from the Ostragradski theorem, but also the ghost and
tachyon problem, can be cast in the requirement that the theory
needs to have an energy functional which is bounded from below.
This condition can certainly not disappear in a consistent low
energy version of a fundamental theory which satisfies it.

The high energy cut-off will be given by some mass
scale, i.e. some Lorentz invariant energy scale of the theory, and
therefore the effective low energy theory should also admit a
Lorentz invariant Lagrangian. Lorentz invariance is not a high
energy phenomenon which can simply be lost at low energies.

What about superluminal motion and causality? We do not want to
require certain properties of the $S$ matrix of the low energy
theory, since the latter may not have a meaningful perturbative
quantum theory; like the 4-Fermi theory, it may not be
renormalizable. Furthermore, one can argue that in cosmology we do
have a preferred frame, the cosmological frame, hence
Lorentz-invariance is broken and we can simply demand that all
superluminal modes of a field propagate forward in cosmic time.
Then no closed signal curves are possible.

But this last argument is very dangerous. Clearly, most solutions
of a Lagrangian theory do break several or most of the symmetries
of the Lagrangian spontaneously. But when applying a Lorentz
transformation to a solution, we produce a new solution that, from
the point of view of the Lagrangian, has the same right of
existence. If some modes of a field propagate with superluminal
speed, this means that their characteristics are spacelike. The
condition that the mode has to travel forward in time with respect
to a certain frame implies that one has to use the retarded
Green's function in this frame. Since spacelike distances have no
frame-independent chronology, for spacelike characteristics this
is a frame-dependent statement. Depending on the frame of
reference, a given mode can represent a normal propagating degree
of freedom, or it can satisfy an elliptic equation, a constraint.

Furthermore, to make sure that the mode propagates forward with
respect to one fixed reference frame, one would have to use
sometimes the retarded, sometimes the advanced and sometimes a
mixture of both functions, depending on the frame of reference. In
a cosmological setting this can be done in a consistent way, but
it is far from clear that such a prescription can be unambiguously
implemented  for generic low energy solutions. Indeed in
Ref.~\cite{adams} a solution is sketched that would not allow
this, so that closed signal curves are again possible.

Therefore, we feel that Lorentz invariant low energy effective
Lagrangians which allow for superluminal propagation of certain
modes, have to be rejected. Nevertheless, this case is not as
clear-cut and there are opposing opinions in the literature,
e.g.~\cite{muki}.

With the advent of the `landscape'~\cite{land}, physicists have
begun to consider anthropic arguments to justify their theory,
whenever it fits the data. Even though the existence of life on
earth is an experimental fact, we consider this argument weak,
nearly tantamount to giving up physics: `Things are like they are
since otherwise we would not be here'. We nevertheless find it
important to inquire also from a purely theoretical point of view,
whether really `anything goes' for effective theories. In the
following sections we shall come back to the basic requirements
which we have outlined in this section.

\section{GENERAL RELATIVISTIC APPROACHES}
\label{s:de}

We give a very brief overview of models for the late-time
acceleration within general relativity, before moving on to the
main topic of modified gravity.

The ``standard" general relativistic interpretation of dark energy
is based on the cosmological constant as vacuum energy:
 \be
G_{\mu\nu}= 8\pi G\left[ T_{\mu\nu}+T^{\text{vac}}_{\mu\nu}
\right],~~T^{\text{vac}}_{\mu\nu}=-{\Lambda_{\rm eff} \over 8\pi
G}\,g_{\mu\nu}\,,
 \ee
where the vacuum energy-momentum tensor is Lorentz invariant. This
approach faces the problem of accounting for the incredibly small
and highly fine-tuned value of the vacuum energy, as summarized in
Eq.~(\ref{v1}).

String theory provides a tantalising possibility in the form of
the ``landscape" of vacua~\cite{land}. There appears to be a vast
number of vacua admitted by string theory, with a broad range of
vacuum energies above and below zero. The idea is that our
observable region of the universe corresponds to a particular
small positive vacuum energy, whereas other regions with greatly
different vacuum energies will look entirely different. This
multitude of regions forms in some sense a ``multiverse". This is
an interesting idea, but it is highly speculative, and it is not
clear how much of it will survive the further development of
string theory and cosmology.

An alternative view of LCDM is the interpretation of $\Lambda$ as
a classical geometric constant~\cite{Padmanabhan:2006cj}, on a par
with Newton's constant $G$. Thus the field equations are
interpreted in the geometrical way,
 \be
G_{\mu\nu}+\Lambda g_{\mu\nu}= 8\pi G T_{\mu\nu}\,.
 \ee
In this approach, the small and fine-tuned value of $\Lambda$ is
no more of a mystery than the host of other fine-tunings in the
constants of nature. For example, more than a 2\% change in the
strength of the strong interaction means that no atoms beyond
hydrogen can form, so that stars and galaxies would not emerge.
But it is not evident whether this distinction between $\La$ and
$\rho_{\text{vac}}$ is really a physical statement, or a purely
theoretical statement that cannot be tested by any experiments.
Furthermore, this classical approach to $\Lambda$ does not evade
the vacuum energy problem -- it simply shifts that problem to
``why does the vacuum not gravitate?" The idea is that particle
physics and quantum gravity will somehow discover a cancellation
or symmetry mechanism to explain why $\rho_{\text{vac}}=0$. This
would be a simpler solution than that indicated by the string
landscape approach, and would evade the disturbing anthropic
aspects of that approach.

Within general relativity, various alternatives to LCDM have been
investigated, in attempt to address the coincidence problem.

\subsection{{\bf DYNAMICAL DARK ENERGY: QUINTESSENCE}}

Here we replace the constant $\Lambda/8\pi G$ by the energy
density of a scalar field $\varphi$, with Lagrangian
 \be
L_\varphi = -\frac{1}{2}g^{\mu\nu}\dd_\mu\varphi \dd_\nu\varphi-
V(\varphi)\,,
 \ee
so that in a cosmological setting,
 \be
&& \rho_\varphi={1\over2}\dot{\varphi}^2+V(\varphi)\,,~ \quad
p_\varphi={1\over2}\dot{\varphi}^2-V(\varphi)\,,\\
&& \ddot\varphi+3H\dot\varphi+V'(\varphi)=0\,,\\
&& H^2 +\frac{K}{a^2} =\frac{8\pi G}{3}\left(\rho_r +\rho_m
+\rho_\varphi\right) \,.
 \ee
The field rolls down its potential and the dark energy density
varies through the history of the universe. ``Tracker" potentials
have been found for which the field energy density follows that of
the dominant matter component. This offers the possibility of
solving or alleviating the fine tuning problem of the resulting
cosmological constant. Although these models are insensitive to
initial conditions, they do require a strong fine-tuning of the
{\em parameters of the Lagrangian} to secure recent dominance of
the field, and hence do not evade the coincidence problem.
An attempt to address the coincidence problem is proposed
in~\cite{wet}, where the transition from the tracker behavior
to dark energy domination is tied to the  neutrino mass.

More
generally, the quintessence potential, somewhat like the inflaton
potential, remains arbitrary, until and unless fundamental physics
selects a potential. There is currently no natural choice of
potential.

In conclusion, there is no compelling reason as yet to choose
quintessence above the LCDM model of dark energy. Quintessence
models do not seem more natural, better motivated or less
contrived than LCDM.  Nevertheless, they are a viable possibility
and computations are straightforward. Therefore, they remain an
interesting target for observations to shoot at~\cite{linder}.

\subsection{{\bf DYNAMICAL DARK ENERGY: MORE GENERAL MODELS}}

It is possible to  couple quintessence  to cold dark matter
without violating current constraints from fifth force
experiments. This could lead to a new approach to the coincidence
problem, since a coupling may provide a less unnatural way to
explain why acceleration kicks in when $\rho_m \sim \rho_{de} $.
In the presence of coupling, the energy conservation equations in
the background become
 \be
\dot\varphi\left[ \ddot\varphi+3H\dot\varphi+V'(\varphi)\right]
&=& Q\,,\\
\dot{\rho}_{\rm dm}+3H\rho_{\rm dm} &=& -Q\,,
 \ee
where $Q$ is the rate of energy exchange. It is relatively simple
to match the geometric data on the background expansion
history~\cite{Amen}. The perturbations show that there is a
momentum transfer as well as an energy transfer. Analysis of the
perturbations typically leads to more stringent constraints, with
some forms of coupling being ruled out by
instabilities~\cite{couppert}.

Another possibility is a scalar field with non-standard kinetic
term in the Lagrangian, for example,
\begin{equation}\label{ns}
{L}_\varphi= F(\varphi,X)-V(\varphi)~\mbox{where}~ X\equiv -{1\over
2}g^{\mu\nu}\partial_\mu\varphi \partial_\nu\varphi\,.
\end{equation}
The standard Lagrangian has $F(\varphi,X)=X$. Some of the
non-standard $F$ models may be ruled out on theoretical grounds.
An example is provided by ``phantom" fields, with negative kinetic
energy density (ghosts), $F(\varphi,X)=-X$. They have $w<-1$, so
that their energy density {\em grows} with expansion. This bizarre
behaviour is reflected in the instability of the quantum vacuum
for phantom fields.

Another example is ``k-essence" fields~\cite{kess}, which have
$F(\varphi,X) = \varphi^{-2}f(X)$. These theories have no ghosts,
and they can produce late-time acceleration. The sound speed of
the field fluctuations for the Lagrangian in Eq.~(\ref{ns}) is
\begin{equation}
{c_s^2}={F_{,X} \over F_{,X}+2XF_{,XX}}\,.
\end{equation}
For a standard Lagrangian, $c_s^2=1$. But for the class of $F$
that produce accelerating k-essence models, it turns out that
there is always an epoch during which $c_s^2>1$, so that these
models may be ruled out according to our causality requirement.
They violate standard causality~\cite{caus}.

For models not ruled out on theoretical grounds, there is the same
general problem as with quintessence, i.e. that no model is better
motivated than LCDM, none is selected by fundamental physics and
any choice of model is more or less arbitrary. Quintessence then
appears to at least have the advantage of simplicity -- although
LCDM has the same advantage over quintessence.

When investigating generic dark energy models we always have to
keep in mind that since both dark energy and dark matter are only
detected gravitationally, we can only measure the total energy
momentum tensor of the dark component,
 \be
T_{\mu\nu}^{\rm dark}=T_{\mu\nu}^{\rm de} + T_{\mu\nu}^{\rm dm} \,
.
 \ee
Hence, if we have no information on the equation of state of dark
energy, there is a degeneracy between the dark energy equation of
state $w(t)$ and $\Om_{\rm dm}$. Without additional assumptions,
we cannot measure either of them by purely gravitational
observations~\cite{martin}. This degeneracy becomes even worse if
we allow for interactions between dark matter and dark energy.

\subsection{{\bf DARK ENERGY AS A NONLINEAR EFFECT FROM
STRUCTURE}}

As structure forms and the matter density perturbation becomes
nonlinear, there are two questions that are posed: (1)~what is the
back-reaction effect of this nonlinear process on the background
cosmology? (2)~how do we perform a covariant and gauge-invariant
averaging over the inhomogeneous universe to arrive at the correct
FRW background? The simplistic answers to these questions are:
(1)~the effect is negligible since it occurs on scales too small
to be cosmologically relevant; (2)~in light of this, the
background is independent of structure formation, i.e., it is the
same as in the linear regime. A quantitative analysis is needed to
fully resolve both issues. However, this is very complicated
because it involves the nonlinear features of general relativity
in an essential way.

There have been claims that these simplistic answers are wrong,
and that, on the contrary, the effects are large enough to mimic
an accelerating universe. This would indeed be a dramatic and
satisfying resolution of the coincidence problem, without the need
for any dark energy field. This issue is discussed
in~\cite{thomas}. Of course, the problem of why the vacuum does
not gravitate would remain.

However, these claims have been disputed, and it is fair to say
that there is as yet no convincing demonstration that acceleration
could emerge naturally from nonlinear effects of structure
formation~\cite{Kolb:2005me}. We should however note that
backreaction/averaging effects could significantly affect our
estimations of cosmological parameters, even if they do not lead
to acceleration~\cite{schwarz}.

It is in principle also possible that the universe around us resembles
more a spherically symmetric but inhomogeneous solution of
Einstein's equation, a Tolman-Bondi-Lema\^\i tre universe, than a
Friedmann-Lema\^\i tre universe. In this case, what appears as
cosmic acceleration to us could perhaps be explained within simple
matter models which only contain dust~\cite{kari}. However, this
would imply that we are situated very close to the centre of a
huge (nearly) spherical structure. Apart from violating the
Copernican principle, this poses another fine tuning problem, and
it also not clear to us whether these models are consistent with all
observations -- not just supernova, but baryon acoustic
oscillations, CMB anisotropies, and weak lensing.

\section{THE MODIFIED GRAVITY APPROACH: DARK GRAVITY}
\label{s:dg}

Late-time acceleration from nonlinear effects of structure
formation is an attempt, within general relativity, to solve the
coincidence problem without a dark energy field. The modified
gravity approach shares the assumption that there is no dark
energy field, but generates the acceleration via ``dark gravity",
i.e. a weakening of gravity on the largest scales, due to a
modification of general relativity itself.

Could the late-time acceleration of the universe be a
gravitational effect? (Note that in general also this does not
remove the problem of why vacuum energy does not gravitate or
is very small.) A historical precedent is provided by attempts
to explain the anomalous precession of Mercury's perihelion by
a ``dark planet´´, named Vulcan. In the end, it was discovered
that a modification to Newtonian gravity was needed.

As we have argued in Section~\ref{s:const}, a consistent
modification of general relativity requires a covariant
formulation of the field equations in the general case, i.e.,
including inhomogeneities and anisotropies. It is not sufficient
to propose ad hoc modifications of the Friedman equation, of the
form
 \be
f(H^2) = {8\pi G\over 3} \rho ~~\mbox{or}~ ~ H^2  = {8\pi G\over
3} g(\rho) \,,
 \ee
for some functions $f$ or $g$. Apart from the fundamental problems
outlined in Section~\ref{s:const}, such a relation allows us to
compute the supernova distance/ redshift relation using this
equation -- but we {\em cannot} compute the density perturbations
without knowing the covariant parent theory that leads to such a
modified Friedman equation. And we also cannot compute the solar
system predictions.

It is very difficult to produce infrared corrections to general
relativity that meet all the minimum requirements:
\begin{itemize}

\item
Theoretical consistency in the sense discussed in
Section~\ref{s:const}.

\item
Late-time acceleration consistent with supernova luminosity
distances, baryon acoustic oscillations and other data that
constrain the expansion history.

\item
A matter-dominated era with an evolution of the scale factor
$a(t)$ that is consistent with the requirements of structure
formation.

\item
Density perturbations that are consistent with the observed growth
factor, matter power spectrum, peculiar velocities, CMB
anisotropies and weak lensing power spectrum.

\item
Stable static spherical solutions for stars, and consistency with
terrestrial and solar system observational constraints.

\item
Consistency with binary pulsar period data.

\end{itemize}

One of the major challenges is to compute the cosmological
perturbations for structure formation in a modified gravity
theory. In general relativity, the perturbations are well
understood. The perturbed metric in Newtonian gauge is
 \be
ds^2=-(1+2\Psi)dt^2+a^2(1+2\Phi)d\vec{x}\,^2, \label{metric}
 \ee
and the metric potentials define two important combinations:
 \be
\Phi_+={1\over2}(\Phi+\Psi)\,,~~~ \Phi_-={1\over2}(\Phi-\Psi)\,.
 \ee
 In the Newtonian limit $\Psi=-\Phi=-\Phi_-$ is the ordinary Newtonian
 potential and $\Phi_+=0$.   The potential $\Phi_+$ is
 sourced by anisotropic stresses. It vanishes if the gravitational field 
is entirely due to non-relativistic matter or a perfect fluid.
The (comoving) matter density perturbation $\Delta=\delta -3aHv$
obeys the Poisson and evolution equations on sub-Hubble scales:
 \be
&& k^2\Phi = 4\pi Ga^2\rho\Delta\,, \\ && \ddot\Delta+2H\dot\Delta
-4\pi G\rho\Delta=0\,.
 \ee
 These equations are exact on all scales, if perturbations are purely
matter ($w=0$) and there are no anisotropic stresses.
On super-Hubble scales (and for adiabatic perturbations, but in
the presence of anisotropic stresses), the evolution of the
perturbations is entirely determined by the
background~\cite{Bertschinger:2006aw} (and the anisotropic stresses
which relate the potentials $\Psi$ and $\Phi$)
 \be
\Phi''-\Psi''-{H'' \over H'}\Phi'-\left({H' \over H}-{H'' \over
H'} \right)\Psi=0\,, \label{bert}
 \ee
where a prime denotes $d/d\ln a$.

The large-angle anisotropies in the CMB temperature encode a
signature of the formation of structure. They are determined by
the propagation of photons along the geodesics of the
perturbed geometry. For adiabatic perturbations one obtains on large
scales the following expression~\cite{mybook} for the temperature
fluctuation in direction $\bf n$:
 \be
\left.{\delta T \over T}({\bf n})\right|_{SW} =
\left[\frac{1}{3}\Psi +\frac{2}{3}H\dot\Phi +\frac{1}{4}\Delta_r +
{\bf V}_b\cdot{\bf n}\right]({\bf x}_\mathrm{dec},t_\mathrm{dec})
-2\int a\dot{\Phi}_-({\bf x}(v),t(v))\,dv\,.
\label{isw}
 \ee
 The integral is along the (unperturbed) trajectory of the light ray with 
affine parameter $v$, from last scattering to today. The position at
decoupling, $\bf x_\mathrm{dec}$, depends on $\bf n$. The same integral
also determines the weak lensing signal, since the deflection angle
is given by (see e.g.~\cite{mybook})
 \be
\vec\alpha = 2 \int \vec\nabla_\perp {\Phi}_-\, dv\,, \label{lens}
 \ee
where $\nabla_\perp$ is the gradient operator in the plane normal to
$\bf n$.

The first term in the square brackets of Eq.~(\ref{isw}) is called
the ordinary Sachs Wolfe effect (OSW), the second term is usually
small since at the time of decoupling the Universe is matter dominated
and this term vanishes in a purely matter dominated Universe. The third term
is responsible for the acoustic peaks in the CMB anisotropy spectrum
and the fourth term is the Doppler term, due to the motion of
the emitting electrons, ${\bf V}_b$ is the baryon velocity field. The
integral is the integrated Sachs Wolfe effect (ISW). It comes from the
fact that the photons are blue shifted when they fall into a gravitational
potential and redshifted when they climb out of it. Hence if the potential
varies during this time, they acquire a net energy shift.

In a modified gravity theory, which we assume to be a metric
theory obeying energy-momentum conservation, Eq.~(\ref{metric})
still holds, and so does the super-Hubble evolution
equation~(\ref{bert}), and the SW and lensing
relations~(\ref{isw}) and (\ref{lens}). But in general
 \be
\Phi_+ \neq0 \,,
 \ee
even in the absence of matter anisotropic stress -- the
modified-gravity effects produce a ``dark" anisotropic stress. In
addition, the Poisson equation and the evolution of density
perturbations will be modified.

\subsection{{\bf $f(R)$ AND SCALAR-TENSOR THEORIES}}

General relativity has a unique status as a theory where gravity
is mediated by a massless spin-2 particle, and the field equations
are second order. Consider modifications to the Einstein-Hilbert
action of the general form
 \be
-\int d^4x\,\sqrt{-g}\,R ~\to~ -\int d^4x\,\sqrt{-g}\,f(R,
R_{\mu\nu}R^{\mu\nu},C_{\mu\nu\alpha\beta}C^{\mu\nu\alpha\beta})\,,
 \ee
where $R_{\mu\nu}$ is the Ricci tensor, $C_{\mu\nu\al\beta} $ is
the Weyl tensor and $f(x_1,x_2,x_3)$ is an arbitrary (at least
three times differentiable) function. Since the curvature tensors
contain second derivatives of the metric, the resulting equations
of motion will in general be fourth order, and gravity is carried
also by massless spin-0 and spin-1 fields. However Ostrogradski's
theorem applies: The usual Hamiltonian formulation of general
relativity leads to six independent metric components $g_{ij}$
which all acquire higher derivative terms. There is actually only
one way out, which is the case $\dd_2f=\dd_3f=0$, i.e., $f$ may
only depend on the Ricci scalar.\footnote{Another possibility
 is the addition of a Gauss Bonnet term, $ \sqrt{-g}f(L_{GB})$,
 where $L_{GB}=R^2-4R_{\mu\nu}R^{\mu\nu}+
 R_{\mu\nu\sigma\rho}R^{\mu\nu\sigma\rho}$. In four
 dimensions $\sqrt{-g}L_{GB}$ contributes only a surface term and
 does not enter the
 equations of motion. However, $ \sqrt{-g}f(L_{GB})$ is non-trivial.
 Such a term also becomes interesting in scalar-tensor theories of
 gravity where one may consider a contribution of the form
 $\sqrt{-g}\phi L_{GB}$ to the Lagrangian.}
The reason is that in the Ricci scalar $R$, only a single component of
the metric contains second derivatives. In this case, the
consequent new degree of freedom can be fixed completely by the
$g_{00}$ constraint, so that the only instability in $f(R)$
theories is the usual one associated with gravitational
collapse~\cite{Woody}.

Therefore, the only acceptable low-energy generalizations of the
Einstein-Hilbert action of general relativity are $f(R)$ theories,
with $f''(R)\neq 0$. The field equations are
 \be
f'(R)R_{\mu\nu}- {1\over2} f(R)g_{\mu\nu}-\left[\nabla_\mu
\nabla_\nu - g_{\mu\nu} \nabla^\alpha \nabla_\alpha \right] f'(R)
=8\pi G T_{\mu\nu}\,,
 \ee
and standard energy-momentum conservation holds:
 \be
\nabla_\nu T^{\mu\nu}=0\,.
 \ee
The trace of the field equations is a wave-like equation for $f'$,
with source term $T=T_\mu\,^\mu$:
 \be
3\nabla^\alpha \nabla_\alpha f'(R)+R f'(R)-2f(R)= 8\pi GT\,.
 \ee
This equation is important for investigating issues of stability
in the theory, and it also implies that Birkhoff's theorem does
not hold.

There has been a revival of interest in $f(R)$ theories due to
their ability to produce late-time
acceleration~\cite{Sotiriou:2008rp}. However, it turns out to be
extremely difficult for this simplified class of modified theories
to pass the observational and theoretical tests. A simple example
of an $f(R)$ model is~\cite{Capozziello:2003tk}
 \be
f(R)=R-{\mu \over R}\,.
 \ee
For $|\mu|\sim H_0^{4}$, this model successfully achieves
late-time acceleration as the $\mu/R$ term starts to dominate. But
the model strongly violates solar system constraints, can have a
strongly non-standard matter era before the late-time
acceleration, and suffers from nonlinear matter
instabilities~\cite{fr}.

\begin{figure*}
\begin{center}
\includegraphics[height=3in,width=2.8in]{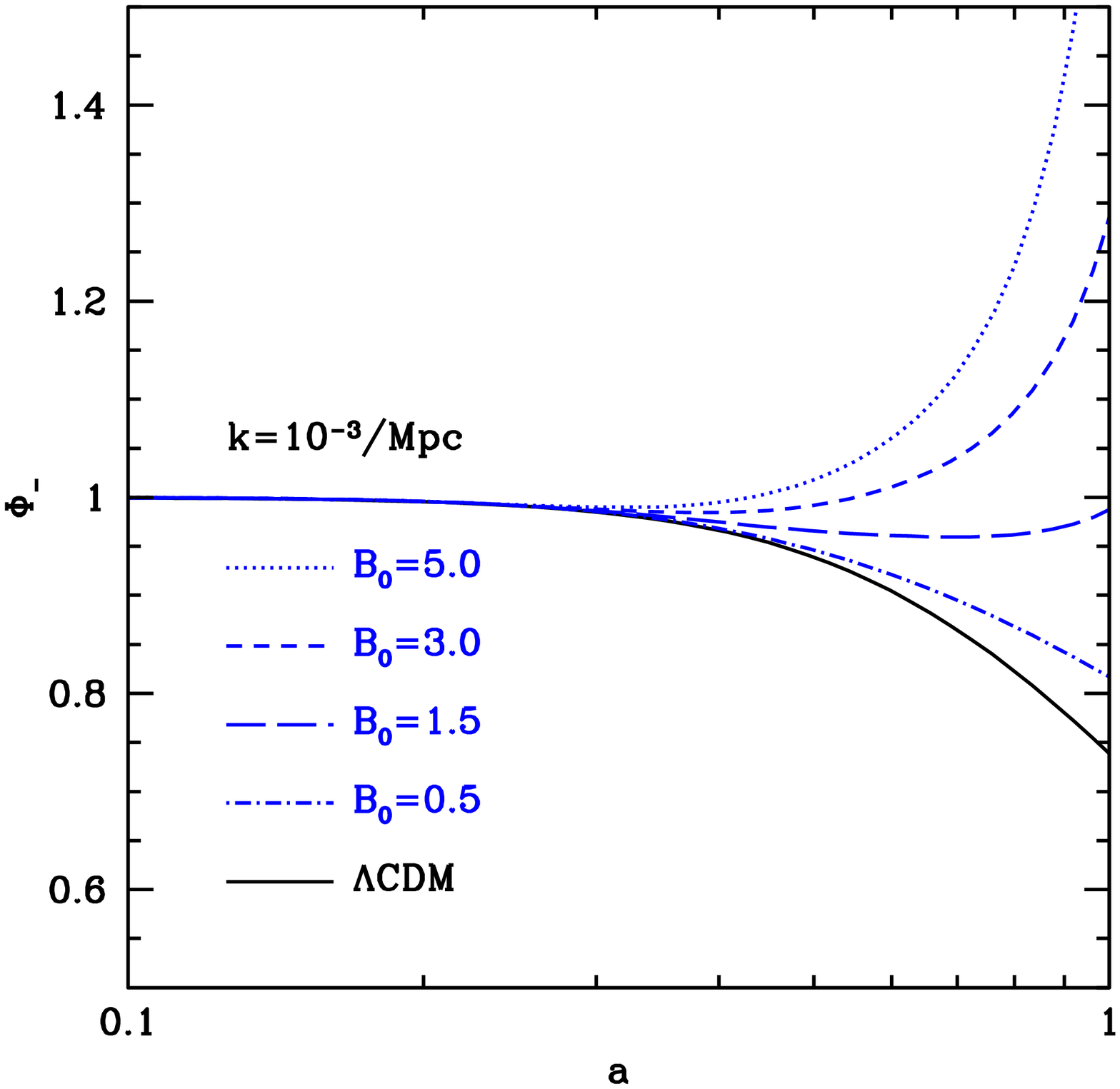}\quad
\includegraphics[height=3in,width=2.8in]{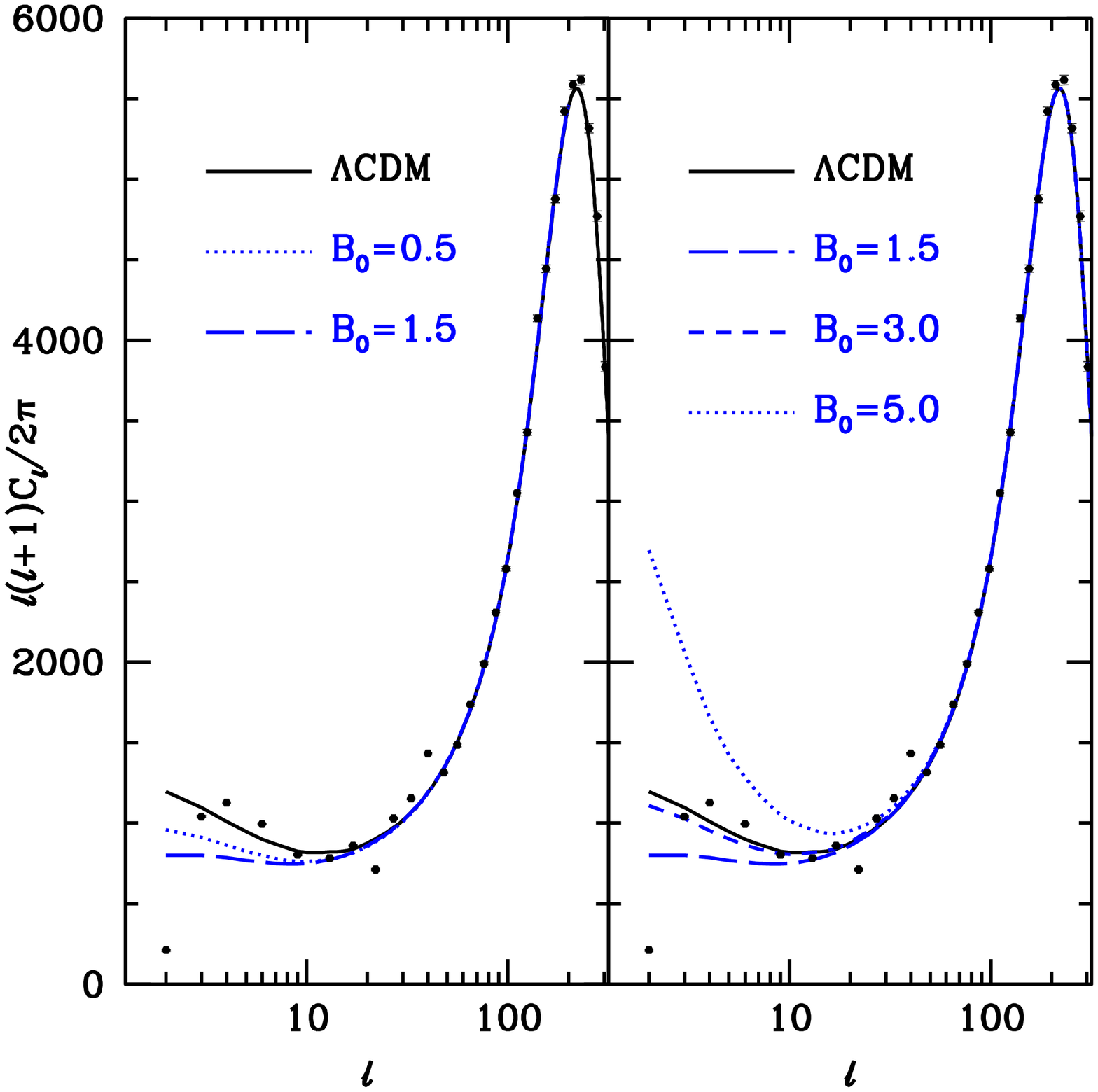}
\caption{{\em Left:} The ISW potential, $(\Phi-\Psi)/2$, for
$f(R)$ models, where the parameter $B_0$ indicates the strength of
deviation from general relativity [see Eq.~(\ref{bpara})].
\\
{\em Right:} The large-angle CMB anisotropies for the models of
the left figure. (For more details see~\cite{Song:2007da}, where
this figure is taken from.)} \label{sph}
\end{center}
\end{figure*}

In $f(R)$ theories, the additional degree of freedom can be
interpreted as a scalar field, and in this sense, $f(R)$ theories
are mathematically equivalent to scalar-tensor theories via
 \be
\psi &\equiv& f'(R)\,,~~
U(\psi) \equiv  -\psi R(\psi) +f(R(\psi)) \,,\\
L &=& -\frac{1}{16\pi G} \sqrt{-g}\left[\psi R +U(\psi)\right].
 \ee
This Lagrangian is the Jordan-frame representation of $f(R)$. It
can be conformally transformed to the Einstein frame, via the
transformation
 \be  \label{e:defphi}
\tilde g_{\mu\nu} = \psi g_{\mu\nu}   \,,~~ \varphi = \sqrt{3\over
4\pi G}\ln\psi ~.
 \ee
In terms of $\tilde g_{\mu\nu}$ and $\varphi$ the Lagrangian then
becomes a standard scalar field Lagrangian,
 \be\label{LphiU}
L = \frac{-1}{16\pi G} \sqrt{-\tilde g}\left[\tilde R
+\frac{1}{2}\tilde g^{\mu\nu}\dd_\mu\varphi \dd_\nu\varphi
+V(\varphi) \right]\,,
 \ee
where
 \be
V(\varphi)= {U(\psi(\varphi)) \over \psi(\varphi)^2} \,.
 \ee
This example shows that modifying gravity (dark gravity) or
modifying the energy momentum tensor (dark energy) can be seen as
a different description of the same physics. Only the coupling of
the scalar field $\varphi$ to ordinary matter, shows that this
theory originates from a scalar-tensor theory of gravity -- and
this non-standard coupling reflects the fact that gravity is also
mediated by a spin-0 degree of freedom, in contrast to general
relativity with a standard scalar field.

The spin-0 field is precisely the cause of the problem with solar
system constraints in most $f(R)$ models, since the requirement of
late-time acceleration leads to a very light mass for the scalar.
The modification to the growth of large-scale structure due to
this light scalar may be kept within observational limits. But on
solar system scales, the coupling of the light scalar to the sun
and planets, induces strong deviations from the weak-field
Newtonian limit of general relativity, in obvious violation of
observations. In terms of the Lagrangian~(\ref{LphiU}) this
scalar has an associated Brans-Dicke parameter that vanishes,
$\omega_{BD}=0$, whereas solar system and binary pulsar data
currently require $\omega_{BD}>40000$.

The only way to evade this problem is to increase the mass of the
scalar near massive objects like the sun, so that the Newtonian
limit can be recovered, while preserving the ultralight mass on
cosmological scales. This ``chameleon" mechanism can be used to
construct models that evade solar system/ binary pulsar
constraints~\cite{newfr}. However the price to pay is that
additional parameters must be introduced, and  the chosen $f(R)$
tends to look unnatural and strongly fine-tuned. An example is
 \be
f(R)=R+\lambda R_0\left[ \left( 1+ {R^2 \over R_0^2}
\right)^{-n}-1\right],
 \ee
where $\lambda, R_0, n $ are positive parameters.

Cosmological perturbations in $f(R)$ theory are well
understood~\cite{Song:2006ej}. The modification to general
relativity produces a dark anisotropic stress
 \be
\Phi_+ \propto {f''(R) \over f'(R)}\,,
 \ee
and deviations from general relativity are conveniently
characterized by the dimensionless parameter
 \be
B={d R/d\ln a \over d \ln H/ d\ln a}\,{f''(R) \over f'(R)}\,.
\label{bpara}
 \ee
If we invoke a chameleon mechanism, then it is possible for these
models to match the observed large-angle CMB anisotropies (see
Fig.~\ref{sph}) and linear matter power
spectrum~\cite{Song:2007da}. However, there may also be fatal
problems with singularities in the strong gravity regime, which
would be incompatible with the existence of neutron
stars~\cite{Kobayashi:2008tq}. These problems appear to arise in
the successful chameleon models, and they are another unintended,
and unexpected, consequence of the scalar degree of freedom, this
time at high energies.

It is possible that an ultraviolet completion of the theory
will cure the high-energy singularity problem. If we assume this
to be the case, then $f(R)$ models that pass the solar system and
late-time acceleration tests are valuable working models for
probing the features of modified gravity theories and for
developing tests of general relativity itself. In order to pursue
this programme, one needs to compute not only the linear
cosmological perturbations and their signature in the growth
factor, the matter power spectrum and the CMB anisotropies -- but
also the weak lensing signal. For this, we need the additional
step of understanding the transition from the linear to the
nonlinear regime. Scalar-tensor behaviour on cosmological scales
relevant to structure formation in the linear regime, must evolve
to Newtonian-like behaviour on small scales in the nonlinear
regime -- otherwise we cannot recover the general relativistic
limit in the solar system. This means that the standard fitting
functions in general relativity cannot be applied, and we require
the development of N-body codes in $f(R)$
theories~\cite{Oyaizu:2008tb}.

More general scalar-tensor theories~\cite{Boisseau:2000pr}, which
may also be motivated via low-energy string theory, have an action
of the form
 \be
-\int d^4x\,\sqrt{-g}\left[ F(\psi)R+{1\over 2}
g^{\mu\nu}\partial_\mu\psi\partial_\nu\psi + U(\psi) \right],
 \ee
where $\psi$ is the spin-0 field supplementing the spin-2
graviton. In the context of late-time acceleration, these models
are also known as ``extended quintessence". Scalar-tensor theories
contain two functions, $F$ and $U$. This additional freedom allows
for greater flexibility in meeting the observational and
theoretical constraints. However, the price we pay is additional
complexity -- and arbitrariness. The $f(R)$ theories have one
arbitrary function, and here there are two, $F(\psi)$ and
$U(\psi)$. There is no preferred choice of these functions from
fundamental theory.

Modifications of the Einstein-Hilbert action, which lead to
fourth-order field equations, either struggle to meet  the minimum
requirements in the simplest cases, or contain more complexity and
arbitrary choices than quintessence models in general relativity.
Therefore, none of these models appears to be a serious competitor
to quintessence in general relativity.

\subsection{{\bf BRANE-WORLD MODELS}}

\begin{figure}[!bth]
\begin{center}
\includegraphics[height=3in,width=3in]{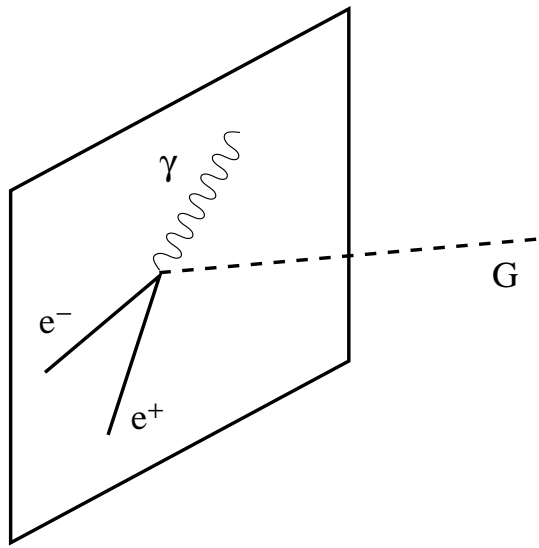}
\caption{The confinement of matter to the brane, while gravity
propagates in the bulk (from~\cite{Cavaglia:2002si}).}
\label{brane}
\end{center}
\end{figure}

Modifications to general relativity within the framework of
quantum gravity are typically ultraviolet corrections that must
arise at high energies in the very early universe or during
collapse to a black hole. The leading candidate for a quantum
gravity theory, string theory, is able to remove the infinities of
quantum field theory and unify the fundamental interactions,
including gravity. But there is a price -- the theory is only
consistent in 9 space dimensions. Branes are extended objects of
higher dimension than strings, and play a fundamental role in the
theory, especially D-branes, on which open strings can end.
Roughly speaking, the endpoints of open strings, which describe
the standard model particles like fermions and gauge bosons, are
attached to branes, while the closed strings of the gravitational
sector can move freely in the higher-dimensional ``bulk"
spacetime. Classically, this is realised via the localization of
matter and radiation fields on the brane, with gravity propagating
in the bulk (see Fig.~\ref{brane}).

The implementation of string theory in cosmology is extremely
difficult, given the complexity of the theory. This motivates the
development of phenomenological models, as an intermediary between
observations and fundamental theory. Brane-world cosmological
models inherit some aspects of string theory, but do not attempt
to impose the full machinery of the theory. Instead,
simplifications are introduced in order to be able to construct
cosmological models that can be used to compute observational
predictions (see~\cite{Maartens:2003tw} for reviews in this
spirit). Cosmological data can then be used to constrain the
brane-world models, and hopefully provide constraints on string
theory, as well as pointers for the further development of string
theory.

It turns out that even the simplest (5D, we effectively
assume that 5 of the extra dimensions in the ``parent" string
theory may be ignored at low energies) brane-world models are
remarkably rich -- and the computation of their cosmological
perturbations is complicated, and in many cases still incomplete.
A key reason for this is that the higher-dimensional graviton
produces a tower of 4-dimensional massive spin-0, spin-1 and
spin-2 modes on the brane, in addition to the standard massless
spin-2 mode on the brane (or in some cases, instead of the
massless spin-2 mode). In the case of some brane models, there are
in addition a massless gravi-scalar and gravi-vector which modify
the dynamics.

Most brane-world models modify general relativity at high
energies. The main examples are those of Randall-Sundrum (RS)
type~\cite{Randall:1999vf}, where a FRW brane is embedded in a
5D anti de Sitter bulk, with curvature radius $\ell$. At low energies
$H\ell \ll 1$, the zero-mode of the graviton dominates on the
brane, and general relativity is recovered to a good
approximation. At high energies, $H\ell\gg 1$, the massive modes
of the graviton dominate over the zero-mode, and gravity on the
brane behaves increasingly five-dimensional. On the brane, the
standard conservation equation holds,
 \be
\dot\rho+3H(\rho+p)=0\,,
  \ee
but the Friedmann equation is modified by an ultraviolet
correction:
 \be\label{mf}
H^2+{K\over a^2} = \frac{8\pi G}{3} \rho\left(1+{2\pi G
\ell^2\over 3}\rho \right) + \frac{\Lambda}{3} \,.
 \ee
The $\rho^2$ term is the ultraviolet correction. At low energies,
this term is negligible, and we recover $H^2+{K/ a^2} \propto
\rho+\Lambda/8\pi G $. At high energies, gravity ``leaks" off the
brane and $H^2\propto \rho^2$. This 5D behaviour means that a
given energy density produces a greater rate of expansion than it
would in general relativity. As a consequence, inflation in the
early universe is modified in interesting
ways~\cite{Maartens:2003tw}.

By contrast, the brane-world model of
Dvali-Gabadadze-Porrati~\cite{Dvali:2000rv} (DGP), which was
generalized to cosmology by Deffayet~\cite{Deffayet:2000uy},
modifies general relativity at {\em low} energies. This model
produces `self-acceleration' of the late-time universe due to a
weakening of gravity at low energies. Like the RS model, the DGP
model is a 5D model with infinite extra dimension.

The action is given by
 \be \label{DGPaction}
{-1\over 16\pi G}\left[ {1\over r_c}\int_{\rm bulk}
d^5x\,\sqrt{-g^{(5)}}\,R^{(5)}+\int_{\rm brane} d^4x\,\sqrt{-g}\,R
\right] \,.
  \ee
The bulk is assumed to be 5D Minkowski spacetime. Unlike the AdS
bulk of the RS model, the Minkowski bulk has infinite volume.
Consequently, there is no normalizable zero-mode of the 4D
graviton in the DGP brane-world. Gravity leaks off the 4D brane
into the bulk at large scales, $r\gg r_c$, where the first term in
the sum~(\ref{DGPaction}) dominates. On small scales, gravity is
effectively bound to the brane and 4D dynamics is recovered to a
good approximation, as the second term dominates. The transition
from 4D to 5D behaviour is governed by the crossover scale $r_c$.
For a Minkowski brane, the weak-field gravitational potential
behaves as
\begin{equation}
\Psi \propto \left\{ \begin{array}{lll} r^{-1} & \mbox{for} & r\ll
r_c
\\ r^{-2} & \mbox{for} & r\gg r_c \end{array}\right.
\end{equation}
On a Friedmann brane, gravity leakage at late times in the
cosmological evolution can initiate acceleration -- not due to any
negative pressure field, but due to the weakening of gravity on
the brane. 4D gravity is recovered at high energy via the lightest
massive modes of the 5D graviton, effectively via an ultra-light
metastable graviton.

\begin{figure}[!bth]
\begin{center}
\includegraphics[height=3in,width=3in]{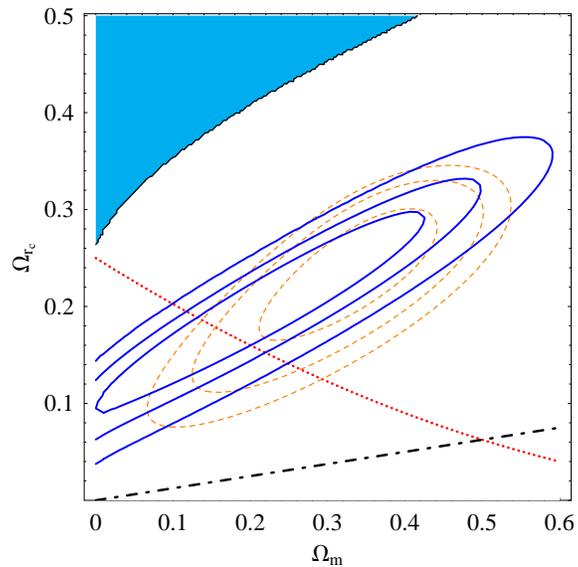}
\caption{The confidence contours for supernova data in the DGP
density parameter plane. The blue (solid) contours are for SNLS
data, and the brown (dashed) contours are for the Gold data. The
red (dotted) curve defines the flat models, the black (dot-dashed)
curve defines zero acceleration today, and the shaded region
contains models without a big bang. (From~\cite{mm}.)}
\label{plane}
\end{center}
\end{figure}

\begin{figure*}
\begin{center}
\includegraphics[width=7.1cm]{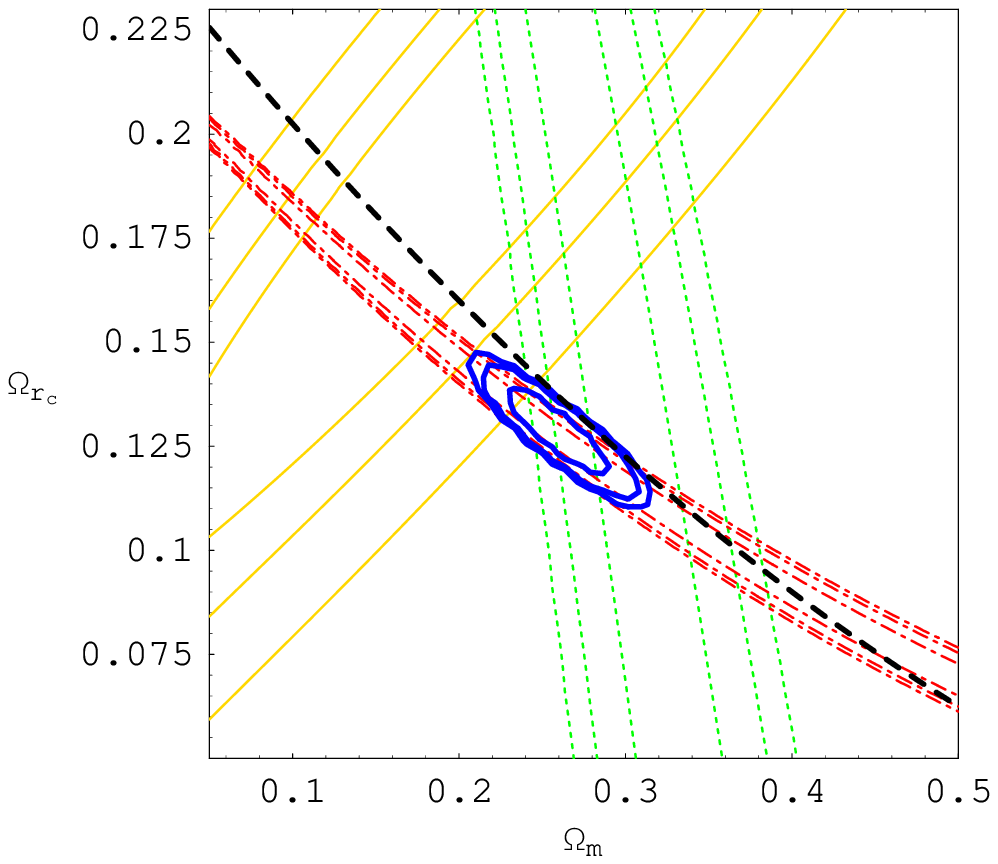}\quad
\includegraphics[width=6.6cm]{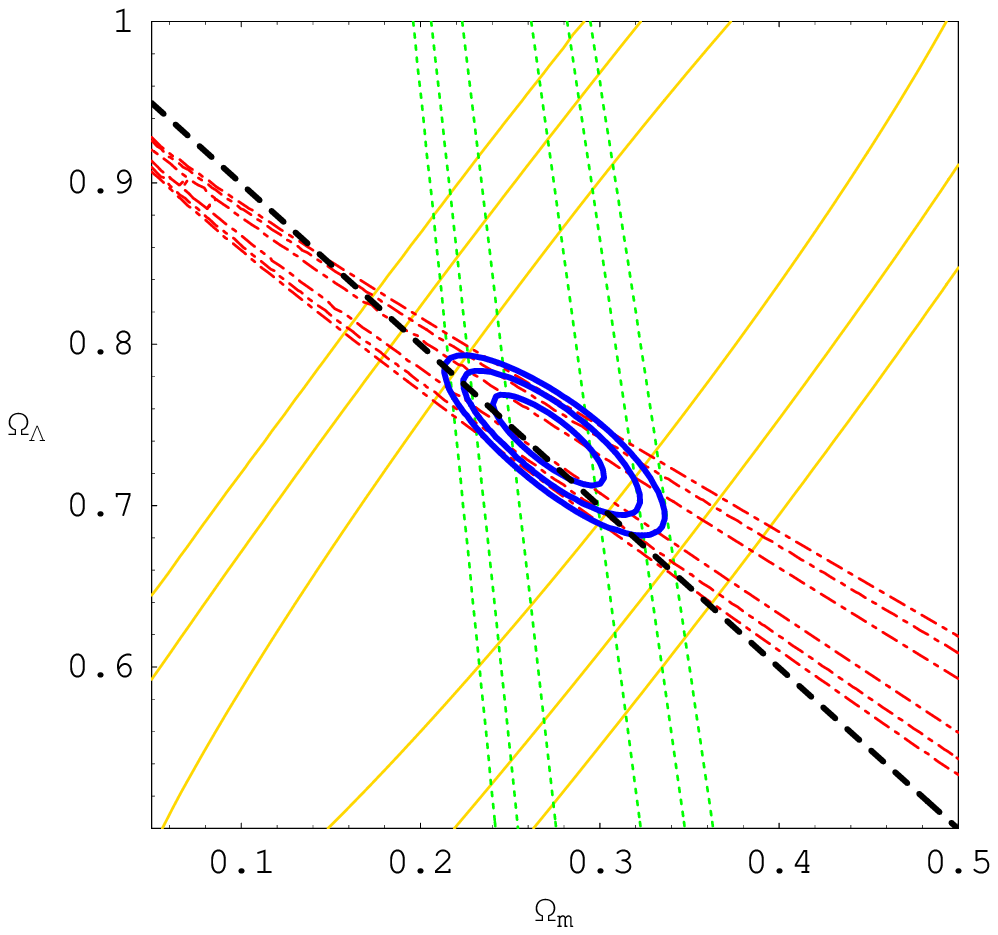}
\caption{Joint constraints [solid thick (blue)] from the SNLS data
[solid thin (yellow)], the BAO peak at $z=0.35 $ [dotted (green)]
and the CMB shift parameter from WMAP3 [dot-dashed (red)]. The
left plot show DGP models , the right plot shows LCDM. The thick
dashed (black) line represents the flat models, $\Omega_K=0$.
(From~\cite{mm}.) }\label{sac}
\end{center}
\end{figure*}

The energy conservation equation remains the same as in general
relativity, but the Friedmann equation is modified:
\begin{eqnarray}
\dot\rho+3H(\rho+p)&=&0\,,\label{ec} \\  H^2+{K\over a^2}-{1 \over
r_c}\sqrt{H^2+{K\over a^2}}&=& {8\pi G \over 3}\rho\,. \label{f}
\end{eqnarray}
To arrive at Eq.~(\ref{f}) we have to take a square root which
implies a choice of sign. As we shall see, the above choice has
the advantage of leading to acceleration but the disadvantage
of the presence of a 'ghost' in this background. It is not
clear whether these facts are related. We shall discuss the
'normal' DGP model, where the opposite sign of the square root
is chosen in the next section.

From Eq.~(\ref{f}) we infer that at early times, i.e., $Hr_c \gg 1$,
the general relativistic Friedman equation is recovered. By contrast, at late
times in an expanding CDM universe, with $\rho\propto a^{-3}\to0$,
we have
\begin{equation}
H\to H_\infty= {1\over r_c}\,,
\end{equation}
so that expansion accelerates and is asymptotically de Sitter. The
above equations imply
 \be
\dot H - {K\over a^2}=-4\pi G\rho\left[1+ {1 \over \sqrt{1+32\pi
Gr_c^2\rho /3}} \right].
 \ee
In order to achieve self-acceleration at late times, we require
 \be
r_c\gtrsim H_0^{-1}\,,
 \ee
since $H_0\lesssim H_\infty$. This is confirmed by fitting
supernova observations, as shown in Fig.~\ref{plane}. The
dimensionless cross-over parameter is defined as
 \be
\Omega_{r_c}={1\over 4(H_0r_c)^2}\,, \label{orc}
 \ee
and the LCDM relation,
 \be
\Omega_m+\Omega_\Lambda+\Omega_K=1\,,
 \ee
is modified to
 \be
\Omega_m+ 2\sqrt{\Omega_{r_c}}\sqrt{1-\Omega_K}+\Omega_K=1\,.
\label{odgp}
 \ee

\begin{figure}[!bth]
\begin{center}
\includegraphics[height=3in,width=3in]{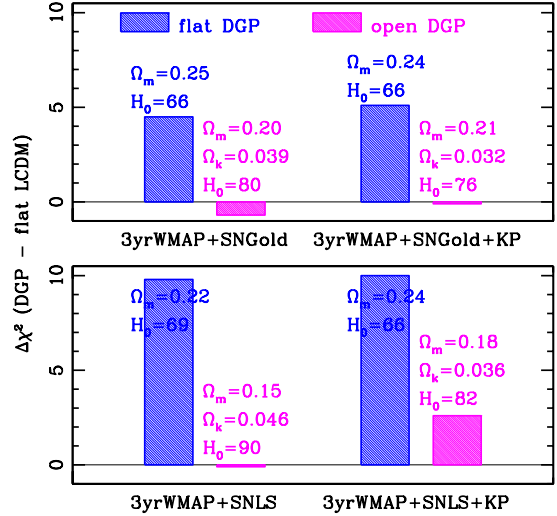}
\caption{The difference in $\chi^2$ between best-fit DGP (flat and
open) and best-fit (flat) LCDM, using SNe, CMB shift and $H_0$ Key
Project data. (From~\cite{Song:2006jk}.)} \label{ssh}
\end{center}
\end{figure}

LCDM and DGP can both account for the supernova observations, with
the fine-tuned values $\Lambda\sim H_0^2$ and $r_c\sim H_0^{-1}$
respectively. When we add further constraints on the expansion
history from the baryon acoustic oscillation peak at $z=0.35 $ and
the CMB shift parameter, the DGP flat models are in strong tension
with data, whereas LCDM models provide a consistent fit. This is
evident in Fig.~\ref{sac}. The open DGP models provide a somewhat
better fit to the geometric data -- essentially because the lower
value of $\Omega_m$ favoured by supernovae reduces the distance to
last scattering and an open geometry is able to extend that
distance. For a combination of SNe, CMB shift and Hubble Key
Project data, the best-fit open DGP also performs better than the
flat DGP~\cite{Song:2006jk}, as shown in Fig.~\ref{ssh}.

Observations based on structure formation provide further
evidence of the difference between DGP and LCDM, since the two
models suppress the growth of density perturbations in different
ways~\cite{Lue:2004rj}. The distance-based observations draw only
upon the background 4D Friedman equation~(\ref{f}) in DGP models
-- and therefore there are quintessence models in general
relativity that can produce precisely the same supernova distances
as DGP. By contrast, structure formation observations require the
5D perturbations in DGP, and one cannot find equivalent
quintessence models~\cite{Koyama:2005kd}. One can find 4D general
relativity models, with dark energy anisotropic stress and
variable sound speed, that  can in principle mimic
DGP~\cite{Kunz:2006ca}. However, these models are highly
unphysical and can be discounted on grounds of theoretical
consistency.

\begin{figure}[t]
\centerline{
\includegraphics[width=9cm]{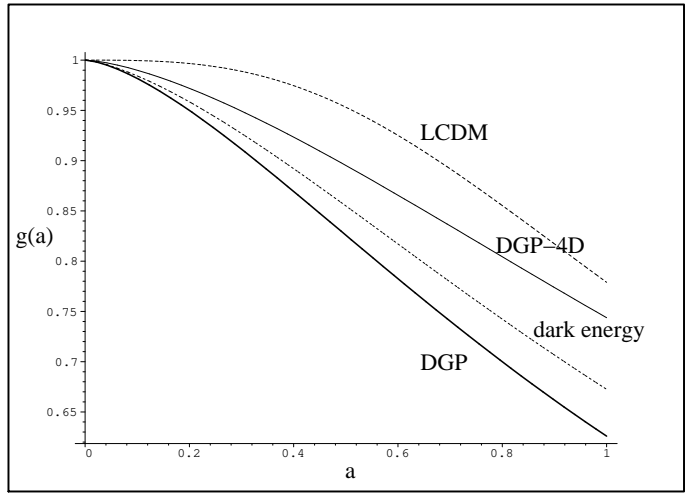}}
\caption{The growth factor $g(a)=\Delta(a)/a$ for LCDM (long
dashed) and DGP (solid, thick), as well as for a dark energy model
with the same expansion history as DGP (short dashed). DGP-4D
(solid, thin) shows the incorrect result in which the 5D effects
are set to zero. (From~\cite{Koyama:2005kd}.)} \label{fig:fig1}
\end{figure}

For LCDM, the analysis of density perturbations is well
understood. For DGP the perturbations are much more subtle and
complicated~\cite{koyama}. Although matter is confined to the 4D
brane, gravity is fundamentally 5D, and the 5D bulk gravitational
field responds to and back-reacts on 4D density perturbations. The
evolution of density perturbations requires an analysis based on
the 5D nature of gravity. In particular, the 5D gravitational
field produces an effective ``dark" anisotropic stress on the 4D
universe. If one neglects this stress and other 5D effects, and
simply treats the perturbations as 4D perturbations with a
modified background Hubble rate -- then as a consequence, the 4D
Bianchi identity on the brane is violated, i.e., $\nabla^\nu
G_{\mu\nu} \neq 0$, and the results are inconsistent. When the 5D
effects are incorporated~\cite{Koyama:2005kd,Cardoso:2007xc}, the
4D Bianchi identity is automatically satisfied. (See
Fig~\ref{fig:fig1}.)

There are three regimes governing structure formation in DGP
models:
\begin{itemize}
\item
On small scales, below the so-called Vainshtein radius (which for
cosmological purposes is roughly the scale of clusters), the
spin-0 scalar degree of freedom becomes strongly coupled, so that
the general relativistic limit is recovered~\cite{Koyama:2007ih}.

\item
On scales relevant for structure formation, i.e. between cluster
scales and the Hubble radius, the spin-0 scalar degree of freedom
produces a scalar-tensor behaviour. A quasi-static approximation
to the 5D perturbations shows that DGP gravity is like a
Brans-Dicke theory with parameter~\cite{Koyama:2005kd}
 \be
\omega_{BD}={3 \over 2}(\beta -1),\label{obd}
 \ee
where
 \be\label{beta}
\beta=1+2H^2r_c\left(H^2+{K \over a^2} \right)^{-1/2}\left[
1+{\dot H \over 3H^2}+{2K \over 3a^2H^2} \right].
 \ee
At late times in an expanding universe, when $Hr_c\gtrsim 1$, it
follows that $\beta<1$, so that $\omega_{BD}<0$. (This signals a
pathology in DGP which is discussed below.)

\item
Although the quasi-static approximation allows us to analytically
solve the 5D wave equation for the bulk degree of freedom, this
approximation breaks down near and beyond the Hubble radius. On
super-horizon scales, 5D gravity effects are dominant, and we need
to solve numerically the partial differential equation governing
the 5D bulk variable~\cite{Cardoso:2007xc}.

\end{itemize}

\begin{figure*}
\begin{center}
\includegraphics[width=2.8in]{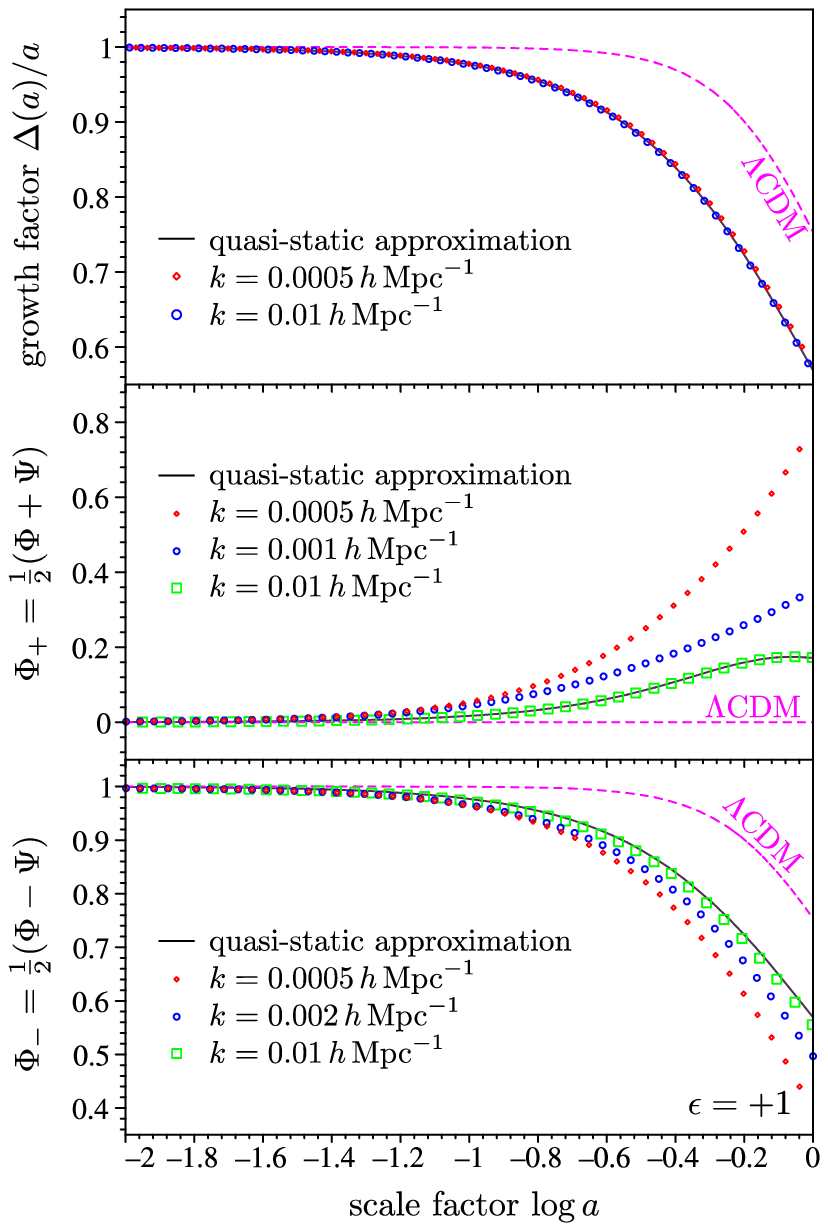}\quad
\includegraphics[width=2.9in]{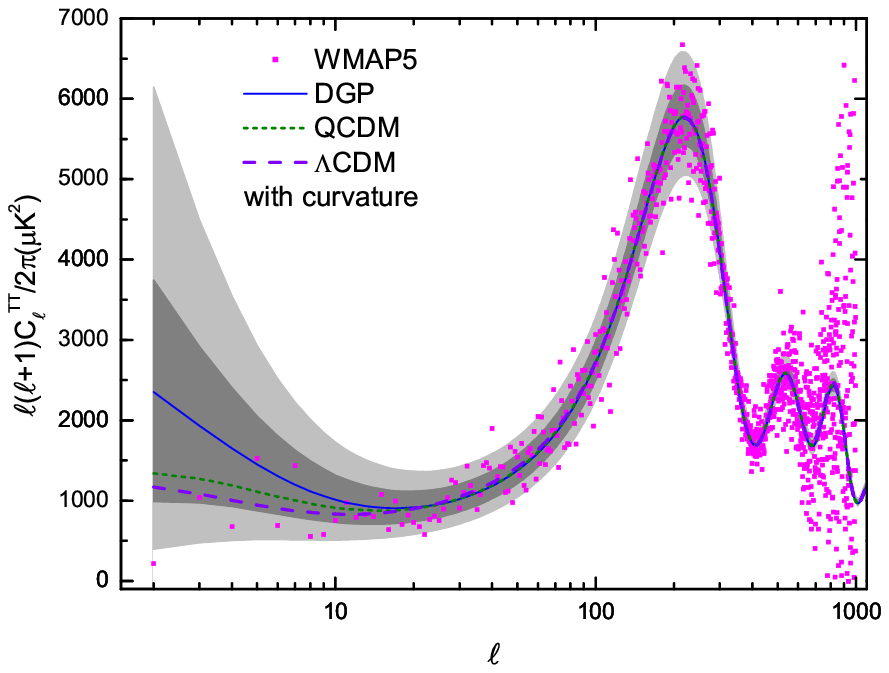}
\caption{{\em Left:} Numerical solutions for DGP density and
metric perturbations, showing also the quasistatic solution, which
is an increasingly poor approximation as the scale is increased.
(From~\cite{Cardoso:2007xc}.) \\ {\em Right:} Constraints on DGP
(the open model that provides a best fit to geometric data) from
CMB anisotropies (WMAP5). The DGP model is the solid curve, QCDM
(short-dashed curve) is the quintessence model with the same
background expansion history as the DGP model, and LCDM is the
dashed curve (a slightly closed model that gives the best fit to
WMAP5, HST and SNLS data). (From~\cite{Fang:2008kc}.)}
\label{card}
\end{center}
\end{figure*}

On sub-horizon scales relevant for linear structure formation, 5D
effects produce a difference between $\Phi$ an $-\Psi$:
 \be
k^2\Phi&=& 4\pi G a^2\left(1-{1\over 3\beta} \right)\rho\Delta\,,
\\ k^2\Psi&=& -4\pi G a^2\left(1+{1\over 3\beta}
\right)\rho\Delta\,,
 \ee
so that there is an effective dark anisotropic stress on the
brane:
 \be
k^2(\Phi+\Psi) = -{8\pi Ga^2 \over 3\beta^2} \rho\Delta\,.
 \ee
The density perturbations evolve as
 \be
\ddot\Delta+2H\dot\Delta -4\pi G\left(1-{1\over 3\beta}
\right)\rho\Delta=0\,.
 \ee
The linear growth factor, $g(a)=\Delta(a)/a$ (i.e., normalized to
the flat CDM case, $\Delta \propto a$), is shown in
Fig.~\ref{fig:fig1}. This shows the dramatic suppression of growth
in DGP relative to LCDM -- from both the background expansion and
the metric perturbations. If we parameterize the growth factor in
the usual way, we can quantify the deviation from general
relativity with smooth dark energy~\cite{Linder:2005in}:
 \be
f:={d\ln\Delta \over d\ln a}=\Omega_m(a)^\gamma\,,~~~~
\gamma\approx \left\{
\begin{array}{lll} 0.55+0.05[1+w(z=1)] & & \text{GR, smooth DE}\\ 0.68 & &
\text{DGP}\end{array} \right.
 \ee

Observational data on the growth factor~\cite{Guzzo:2008ac} are
not yet precise enough to provide meaningful constraints on the
DGP model. Instead, we can look at the large-angle anisotropies of
the CMB, i.e. the ISW effect. This requires a treatment of
perturbations near and beyond the horizon scale. The full
numerical solution has been given by~\cite{Cardoso:2007xc}, and is
illustrated in Fig.~\ref{card}. The CMB anisotropies are also
shown in  Fig.~\ref{card}, as computed in~\cite{Fang:2008kc} using
a scaling approximation to the super-Hubble
modes~\cite{Sawicki:2006jj} (the accuracy of the scaling ansatz is
verified by the numerical results~\cite{Cardoso:2007xc}).

It is evident from Fig.~\ref{card} that the DGP model which
provides a best fit to the geometric data (see Fig.~\ref{ssh}), is
in serious tension with the WMAP5 data on large scales. The
problem arises form the large deviation of $\Phi_-=(\Phi-\Psi)/2$
in the DGP model from the LCDM model. This deviation, i.e. a
stronger decay of $\Phi_-$, leads to an over-strong ISW effect
[see Eq.~(\ref{isw})], in tension with WMAP5 observations.

As a result of the combined observations of background expansion
history and large-angle CMB anisotropies, the DGP model provides a
worse fit to the data than LCDM at about the 5$\sigma$
level~\cite{Fang:2008kc}. Effectively, the DGP model is ruled out
by observations in comparison with the LCDM model.

In addition to the severe problems posed by cosmological
observations, a problem of theoretical consistency is posed by the
fact that the late-time asymptotic de Sitter solution in DGP
cosmological models has a ghost. The ghost is signaled by the
negative Brans-Dicke parameter in the effective theory that
approximates the DGP on cosmological sub-horizon scales:
 \be
\omega_{BD}<0\,.
 \ee
The existence of the ghost is confirmed by detailed analysis of
the 5D perturbations in the de Sitter
limit~\cite{Gorbunov:2005zk,Charmousis:2006pn}. The DGP ghost is a
ghost mode in the scalar sector of the gravitational field --
which is more serious than the ghost in a phantom scalar field. It
effectively rules out the DGP, since it is hard to see how an
ultraviolet completion of the DGP can cure the {\em infrared}
ghost problem..

\subsection{{\bf DEGRAVITATION AND NORMAL DGP}}

The self-accelerating DGP is effectively ruled out as a
cosmological model by observations and by the problem of the ghost
in the gravitational sector. Indeed, it may be the case that
self-acceleration comes with the price of ghost states. An
alternative idea is that massive-graviton theories (like the DGP)
may lead to {\em degravitation}~\cite{deRham:2007xp}, i.e., the
feature that the vacuum energy (cosmological constant), does not
gravitate at the level expected [as in Eq.~(\ref{v1})], and
possibly not at all.

To achieve a reduction of gravitation on very large scales,
degravitation, Newton's constant is promoted to a 'high-pass filter'
and Einstein's equations are modified to
\be
G^{-1}(L^2\Box)G_{\mu\nu} =8\pi T_{\mu\nu} ~.
\ee
We want $G(L^2\Box)$ to act as a high pass filter: for scales smaller
than $L$ it is constant while scales much larger than $L$ are filtered out,
degravitated. For this to work, $G^{-1}$  must contain inverse
powers of $\Box$, hence it must be non-local. Furthermore, this equation
cannot describe a massless spin 2 graviton with only two degrees of
freedom, but it leads, at the linear level to  massive gravitons with mass
$1/L$ or a superposition (spectral density) of massive gravitons. These are
known to carry three
additional polarizations two of helicity 1 and one helicity 0 state. The
latter couples to the trace of the energy momentum tensor and remains
present also in the zero-mass limit, the well known
van Dam-Veltman-Zakharov discontinuity of massive gravity~\cite{vDVZ}.
This problem might be solved on small scales, where the extra polarizations
become strongly coupled due to non-linear self interactions~\cite{Vainst}.
One can show that in regions where the curvature exceeds $L^{-2}$, the
extra polarizations are suppressed by powers of $L$ and we recover
ordinary spin-2 gravity.

Contrary to the models discussed so far, these theories can in principle 
address the cosmological constant problem: the cosmological constant is not
necessarily small, but we cannot see it in gravitational experiments since
it is (nearly) degravitated. On the other hand, the problem of the present 
cosmological acceleration is not addressed.

\begin{figure*}
\begin{center}
\includegraphics[width=2.8in]{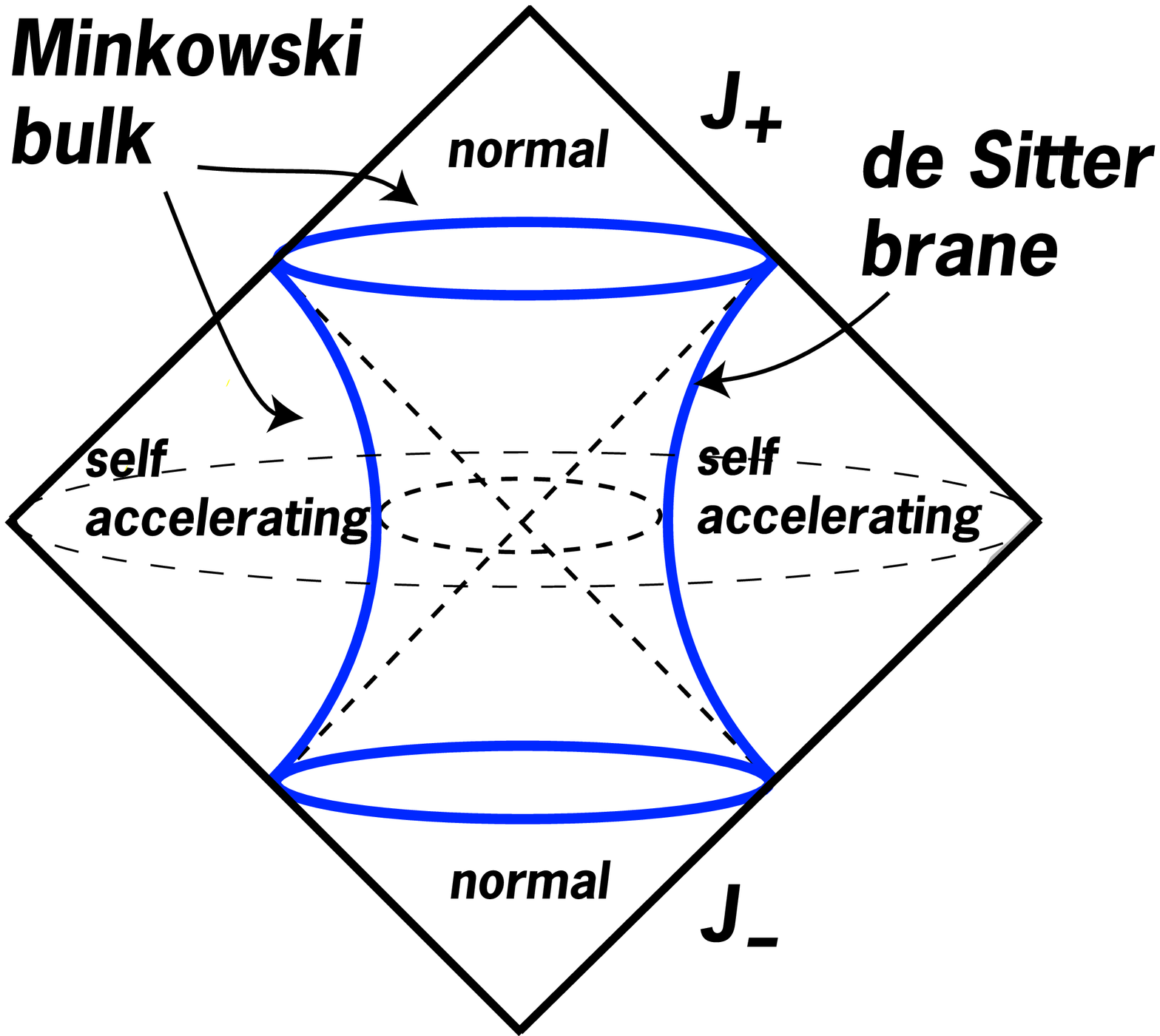} \quad
\includegraphics[width=2.8in]{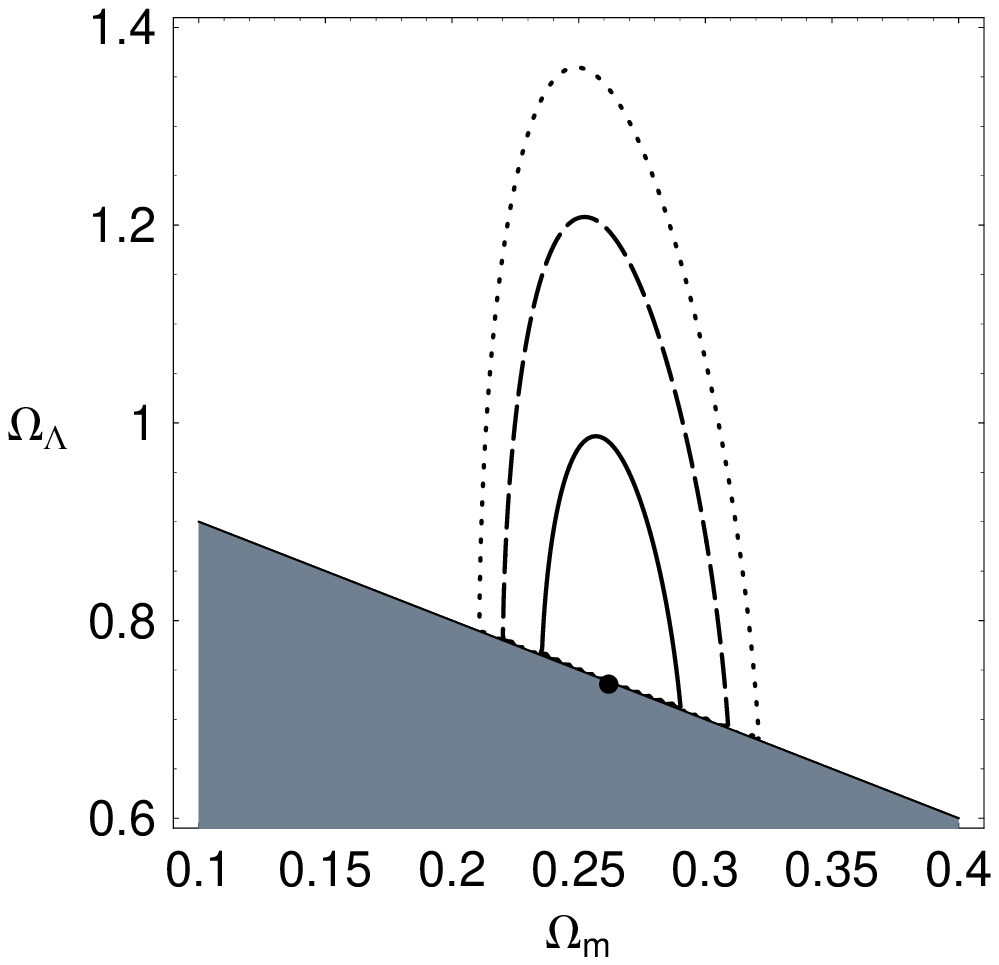}
\caption{{\em Left:} The embedding of the self-accelerating and
normal branches of the DGP brane in a Minkowski bulk.
(From~\cite{Charmousis:2006pn}.)
\\ {\em Right:} Joint constraints on normal DGP (flat, $K=0$)
from SNLS, CMB shift
(WMAP3) and BAO ($z=0.35$) data. The best-fit is the solid point,
and is indistinguishable from the LCDM limit. The shaded region is
unphysical and its upper boundary represents flat LCDM models.
(From~\cite{Lazkoz:2006gp}.)} \label{ndgp}
\end{center}
\end{figure*}

Apart from a simple massive graviton, the simplest example of degravitation 
is provided by the so-called
``normal" (i.e., non-self-accelerating and ghost free) branch of the
DGP~\cite{Sahni:2002dx}, which arises from a different embedding
of the DGP brane in the Minkowski bulk (see Fig.~\ref{ndgp}). In
the background dynamics, this amounts to a replacement $r_c \to
-r_c $ in Eq.~(\ref{f}) -- and there is no longer
late-time self-acceleration. It is therefore necessary to include
a $\Lambda$ term in order to accelerate the late universe:
\begin{eqnarray}
H^2+{K\over a^2}+{1 \over r_c}\sqrt{H^2+{K\over a^2}}= {8\pi G
\over 3}\rho +{\Lambda \over3}\,. \label{fn}
\end{eqnarray}
(Normal DGP models with a quintessence field have also been
investigated~\cite{Chimento:2006ac}.) Using the dimensionless
crossover parameter defined in Eq.~(\ref{orc}), the densities are
related at the present time by
 \be
\sqrt{1-\Omega_K}=-\sqrt{\Omega_{r_c}} +\sqrt{\Omega_{r_c} +
\Omega_m + \Omega_\Lambda }\,, \label{ondgp}
 \ee
which can be compared with the self-accelerating DGP
relation~(\ref{odgp}).

\begin{figure*}
\begin{center}
\includegraphics[width=2.8in]{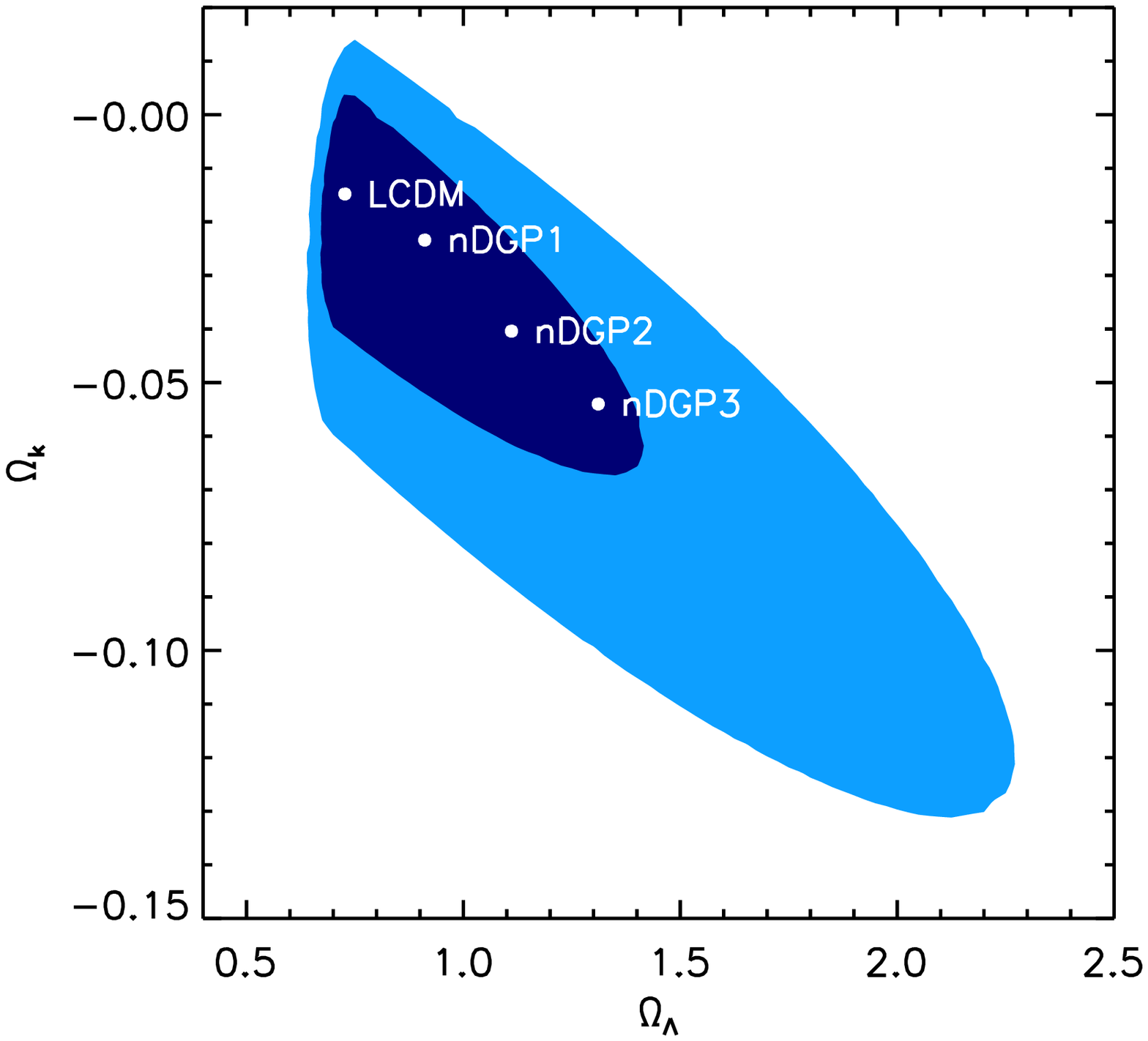}\quad
\includegraphics[width=2.8in]{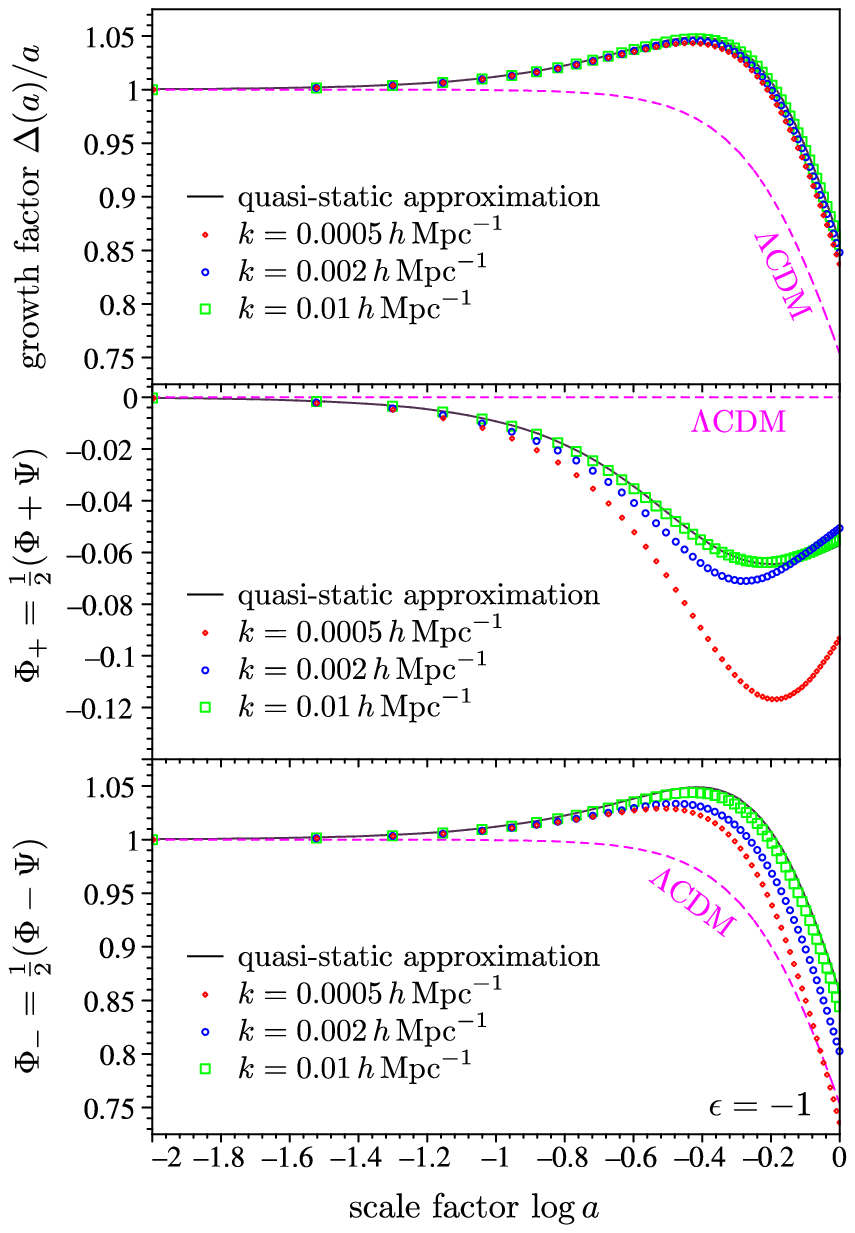}
\caption{{\em Left:} Joint constraints on normal DGP from SNe
Gold, CMB shift (WMAP3) and $H_0$ data in the projected
curvature-$\Lambda$ plane, after marginalizing over other
parameters. The best-fits are the solid points, corresponding to
different values of $\Omega_m$.
(From~\cite{Giannantonio:2008qr}.)\\
{\em Right:} Numerical solutions for the normal DGP density and
metric perturbations, showing also the quasistatic solution, which
is an increasingly poor approximation as the scale is increased.
Compare with the self-accelerating DGP case in Fig.~\ref{card}.
(From~\cite{Cardoso:2007xc}.) } \label{gsk}
\end{center}
\end{figure*}

The degravitation feature of normal DGP is that $\Lambda$ is
effectively screened by 5D gravity effects. This follows from
rewriting the modified Friedmann equation~(\ref{fn}) in standard
general relativistic form, with
 \be
\Lambda_{\rm eff}=\Lambda -{3 \over r_c} \sqrt{H^2+{K\over a^2}} <
\Lambda \,.
 \ee
Thus 5D gravity in normal DGP can in principle reduce the bare
vacuum energy significantly. However, figure~\ref{ndgp} shows that best-fit
flat models, using geometric data, only admit insignificant
screening~\cite{Lazkoz:2006gp}.  The closed models provide
a better fit to the data~\cite{Giannantonio:2008qr}, and can allow
a bare vacuum energy term with $\Omega_\Lambda >1$, as shown in
Fig.~\ref{gsk}. This does not address the fundamental problem of
the smallness of $\Omega_\Lambda$, but it is nevertheless an
interesting feature. We can define an effective equation of state
parameter via
 \be
\dot{\Lambda}_{\rm eff}+3H(1+w_{\rm eff}) \Lambda_{\rm eff}=0\,.
 \ee
At the present time (setting $K=0$ for simplicity),
 \be
w_{\rm eff,0}=-1 -{(\Omega_m+\Omega_\Lambda-1)\Omega_m \over
(1-\Omega_m) (\Omega_m+\Omega_\Lambda+1)}<-1\,,
 \ee
where the inequality holds since $\Omega_m<1$. This reveals
another important property of the normal DGP model: effective
phantom behaviour of the recent expansion history. This is
achieved without any pathological phantom field (similar to what
can be done in scalar-tensor theories~\cite{Boisseau:2000pr}).
Furthermore, there is no ``big rip" singularity in the future
associated with this phantom acceleration, unlike the situation
that typically arises with phantom fields. The phantom behaviour
in the normal DGP model is also not associated with any ghost
problem -- indeed, the normal DGP branch is free of the ghost that
plagues the self-accelerating DGP~\cite{Charmousis:2006pn}.

Perturbations in the normal branch have the same structure as
those in the self-accelerating branch, with the same regimes --
i.e. below the Vainshtein radius (recovering a GR limit), up to the
Hubble radius (Brans-Dicke behaviour), and beyond the Hubble
radius (strongly 5D behaviour). The quasistatic approximation and
the numerical integrations can be simply repeated with the
replacement $r_c \to -r_c$ (and the addition of $\Lambda$ to the
background). In the sub-Hubble regime, the effective Brans-Dicke
parameter is still given by Eqs.~(\ref{obd}) and (\ref{beta}), but
now we have $\omega_{BD}>0$ -- and this is consistent with the
absence of a ghost. Furthermore, a positive Brans-Dicke parameter
signals an extra positive contribution to structure formation from
the scalar degree of freedom, so that there is {\em less}
suppression of structure formation than in LCDM -- the reverse of
what happens in the self-accelerating DGP. This is confirmed by
computations, as illustrated in Fig.~\ref{gsk}.

The closed normal DGP models fit the background expansion data
reasonably well, as shown in Fig.~\ref{gsk}. The key remaining
question is how well do these models fit the large-angle CMB
anisotropies, which is yet to be computed at the time of writing.
The derivative of the ISW potential $\dot{\Phi}_-$ can be seen in
Fig.~\ref{gsk}, and it is evident that the ISW contribution is
negative relative to LCDM at high redshifts, and goes through zero
at some redshift before becoming positive. This distinctive
behaviour may be contrasted with the behaviour in $f(R)$ models
(see Fig.~\ref{sph}): both types of model lead to less suppression
of structure than LCDM, but they produce different ISW
effects. However, in the limit $r_r\to \infty$, normal DGP tends to ordinary
LCDM, hence observations which fit LCDM will always just provide a
lower limit for $r_c$.

\section{CONCLUSION}
\label{s:con}

The evidence for a late-time acceleration of the universe
continues to mount, as the number of experiments and the quality
of data grow. This revolutionary discovery by observational
cosmology, confronts theoretical cosmology with a major crisis --
how to explain the origin of the acceleration. The core of this
problem may be ``handed over" to particle physics, since we
require at the most fundamental level, an explanation for why the
vacuum energy either has an incredibly small and fine-tuned value,
or is exactly zero. Both options violently disagree with naive
estimates of the vacuum energy.

If one accepts that the vacuum energy is indeed nonzero, then the
dark energy is described by $\Lambda$, and the LCDM model is the
best current model. The cosmological model requires completion via
developments in particle physics that will explain the value of
the vacuum energy. In many ways, this is the best that we can do
currently, since the alternatives to LCDM, within and beyond
general relativity, do not resolve the vacuum energy crisis, and
furthermore have no convincing theoretical motivation. None of the
contenders so far appears any better than LCDM, and it is fair to
say that at the theoretical level, there is as yet no serious
challenger to LCDM. One consequence of this is the need to develop
better observational tests of LCDM, which could in principle rule
it out, e.g. by showing, to some acceptable level of statistical
confidence, that $w\neq -1$. However, observations are still quite
far from the necessary precision for this.

It remains necessary and worthwhile to continue investigating
alternative dark energy and dark gravity models, in order better
to understand the space of possibilities, the variety of
cosmological properties, and the observational strategies needed
to distinguish them. The lack of any consistent and compelling
theoretical model means that we need to keep exploring
alternatives -- and also to keep challenging the validity of
general relativity itself on cosmological scales.

We have focused in this chapter on two of the simplest
infrared-modified gravity models: the $f(R)$ models (the simplest
scalar-tensor models), and the DGP models (the simplest
brane-world models). In both types of model, the new scalar degree
of freedom introduces severe difficulties at theoretical and
observational levels. Strictly speaking, the $f(R)$ models are
probably ruled out by the presence of singularities that exclude
neutron stars (even if they can match all cosmological
observations, including weak lensing). And the DGP models are
likely ruled out by the appearance of a ghost in the asymptotic de
Sitter state -- as well as by a combination of geometric and
structure-formation data.

Nevertheless, the intensive investigation of $f(R)$ and DGP models
has left an important legacy -- in a deeper understanding of
\begin{itemize}
\item
the interplay between gravity and expansion history and structure;
\item
the relation between cosmological and local observational
constraints;
\item
the special properties of general relativity itself;
\item
the techniques needed to distinguish different candidate models,
and the limitations and degeneracies within those techniques;
\item
the development of tests that can probe the validity of general
relativity itself on cosmological scales, independent of any
particular alternative model.
\end{itemize}

The last point is one of the most important by-products of the
investigation of modified gravity models. It involves a careful
analysis of the web of consistency relations that link the
background expansion to the evolution of
perturbations~\cite{testgr}, and opens up the real prospect of
testing general general relativity well beyond the solar system
and its neighbourhood.
\vspace{1cm}

{\bf Acknowledgments:}\\
We thank Camille Bonvin, Chiara Caprini, Kazuya Koyama, Martin
Kunz, Sanjeev Seahra and Norbert Straumann for stimulating and
illuminating discussions. This work is supported by the Swiss
National Science Foundation and the UK STFC.

\end{document}